# Superior ferroelectricity and nonlinear optical response in a hybrid germanium iodide hexagonal perovskite

Kun Ding[1,2,#], Haoshen Ye[3,#], Changyuan Su[1,#], Yu-An Xiong[1], Guowei Du[1], Yu-Meng You[1], Zhi-Xu Zhang[1,2], Shuai Dong[3],Yi Zhang[2] and Da-Wei Fu[2]

[1] Jiangsu Key Laboratory for Science and Applications of Molecular Ferroelectrics, Southeast University, Nanjing, 211189, China

[2] Institute for Science and Applications of Molecular Ferroelectrics, Key Laboratory of the Ministry of Education for Advanced Catalysis Materials, Zhejiang Normal University, Jinhua, 321019, China

[3] Key Laboratory of Quantum Materials and Devices of Ministry of Education, School of Physics, Southeast University, Nanjing, 211189, China

[#] These authors contributed equally: Kun Ding, Haoshen Ye, Changyuan Su

Correspondence and requests for materials should be addressed to Z.-X. Z. (email: zhixu@seu.edu.cn); S.D. (email: sdong@seu.edu.cn); Y.Z. (email: yizhang1980@seu.edu.cn) or to D.-W. F. (email: dawei@seu.edu.cn)

## Abstract

Abundant chemical diversity and structural tunability make organic-inorganic hybrid perovskites (OIHPs) a rich ore for ferroelectrics. However, compared with their inorganic counterparts such as $BaTiO_3$, their ferroelectric key properties, including large spontaneous polarization ($P_s$), low coercive field ($E_c$), and strong second harmonic generation (SHG) response, have long been great challenges, which hinder their commercial applications. Here, a quasi-one-dimensional OIHP $DMAGeI_3$ (DMA= Dimethylamine) is reported, with notable ferroelectric attributes at room temperature: a large $P_s$ of 24.14 μC/cm$^2$ (on a par with $BaTiO_3$), a low $E_c$ below 2.2 kV/cm, and the strongest SHG intensity in OIHP family (about 12 times of $KH_2PO_4$ (KDP)). Revealed by the first-principles calculations, its large $P_s$ originates from the synergistic effects of the stereochemically active $4s^2$ lone pair of $Ge^{2+}$ and the ordering of organic cations, and its low kinetic energy barrier of small DMA cations results in a low $E_c$. Our work brings the comprehensive ferroelectric performances of OIHPs to a comparable level with commercial inorganic ferroelectric perovskites.

## Introduction

Ferroelectrics with switchable spontaneous electrical polarizations in response to electric fields, having demonstrated great scientific and application values in data storage, sensors, optoelectronics, and versatile micro-electro-mechanical systems (MEMSs)[1-7]. As the most important properties of ferroelectricity: spontaneous polarization ($P_s$), coercive electric field ($E_c$), and Curie temperature ($T_C$), vitally determine the comprehensive performances of ferroelectric materials and their commercial value for applications[8-11]. Inorganic oxide ferroelectrics, e.g. perovskite $BaTiO_3$ and $Pb(Zr, Ti)O_3$, have long been the successful mainstream of commercial applications, thanks to their high $T_C$, large $P_s$, small $E_c$, as well as chemical stability[12-15]. However, their rigid structures, non-passive surfaces, and incompatibility with silicon industry, becomes more and more challenging for next-generation devices, such as nanoscale MEMSs and flexible electronics.

With the requirements from flexible applications, organic-inorganic hybrid perovskite (OIHP) ferroelectrics have emerged as a promising branch, by taking the advantage of low-cost manufacture, environmentally friendly, ease of fabrication, and excellent film formation[16-21]. OIHP ferroelectrics own rich structural diversity[22-28], including three-dimensional (3D) $ABX_3$-type (pseudo-)cubic structures[29,30], two-dimensional (2D) layered structures[31,32], and even quasi-one-dimensional (1D) $ABX_3$-type hexagonal structures[33]. Different with 3D and 2D OIHPs whose structures are limited by many issues such as tolerance factor, 1D OIHPs do not have strict selections on metal halide anions and organic cations regarding their sizes, shapes, or valences. Thus 1D OIHPs are endowed with great structural freedom to engineer high-performance ferroelectrics with intriguing functionalities. Indeed, some 1D OIHP ferroelectrics have been synthesized and exhibit interesting properties such as large piezoelectric response[34], multi-step nonlinear optical switches[35], photoluminescence, magnetism[36], and circularly polarized luminescence[37]. In particular, two 1D OIHP ferroelectrics $[Me_3NCH_2Cl]MCl_3$ ($M$=Mn and Cd) have been reported to show large piezoelectric response on par with $BaTiO_3$[38]. Even though, the comprehensive performances of available OIHP ferroelectrics remain incompetent when competing with conventional ferroelectric oxides. The most serious problems are their small $P_s$ (typically <10 μC/cm$^2$) and relatively high $E_c$ (~100 kV/cm). For comparison, the $P_s$ of $BaTiO_3$ reaches 26 μC/cm$^2$ and its typical $E_c$ is ~10 kV/cm[39].

Screening and modifying organic cations (imidazole, pyrrolidine, quinuclidine, etc.) are efficient strategies to improve the material performance[40]. Because the polarization switching of OIHPs is generally related to the ordering reorientation of organic cations, it appears that small organic cations may be beneficial in reducing the coercive field. Besides, the inorganic framework of $BX_6$ octahedra is also with remarkably compositional diversity. But their contributions to polarization have mostly been omitted since those octahedra formed by common transition metals (i.e., Cd, Cr, Mn, Cu) are generally inactive to ferroelectricity.

Inspired by the aforementioned ideas (Supplementary Fig. 1), we used the small $DMA^+$ cation and germanium halide to synthesize a series of OIHP $DMAGeX_3$ ($X$=Cl, Br, I) (Supplementary Note 1). Remarkably, OIHP $DMAGeI_3$ exhibits excellent ferroelectricity with $T_C$=363 K, a high $P_s$=24.14 μC/cm$^2$, and a low $E_c$=0.8-2.2 kV/cm at room temperature. To our best knowledge, such an experimental $P_s$ is higher than those of most reported pure-organic and inorganic-organic hybrid ferroelectrics, and rivals the popular ferroelectric oxide $BaTiO_3$. And equally importantly, its $E_c$ value is one order of magnitude lower than that of recently reported inorganic perovskite $CsGeI_3$ (~40 kV/cm)[41], and two orders of magnitude lower than those of PVDF (~500 kV/cm) and its copolymers[39]. Moreover, it owns a large SHG response, whose intensity is more than ten times stronger than that of $KH_2PO_4$ (KDP, a commonly used standard system). Our finding brings the comprehensive performances of OHIP ferroelectrics to a comparable level with typical inorganic ferroelectric perovskites.

## Results

### Structural analysis of crystal

Yellow and transparent rod-shaped single crystals of $DMAGeI_3$ in size of 2×2×16 $mm^3$ (Fig. 1a) were prepared by hydrothermal method (Supplementary Note 1). The phase purity and thermal stabilities of grown crystals were identified by powder X-ray diffraction (XRD) and thermogravimetric analysis (TGA), which shows good environmental and thermal stability (Supplementary Figs. 2, 3). Phase transition of $DMAGeI_3$ was characterized using differential scanning calorimetry (DSC) measurements. As shown in Supplementary Fig. 4, the DSC curves show a pair of thermal anomalies peaked at 359 K and 363 K in cooling/heating runs, respectively. For convenience, we defined the phase above $T_C$ of 363 K as the high-temperature phase (HTP) and the phase below $T_C$ as the low-temperature phase (LTP). Crystal structures in LTP and HTP were determined by variable-temperature single-crystal X-ray diffraction in the range of 223-373 K (Supplementary Note 2 & Table 1).

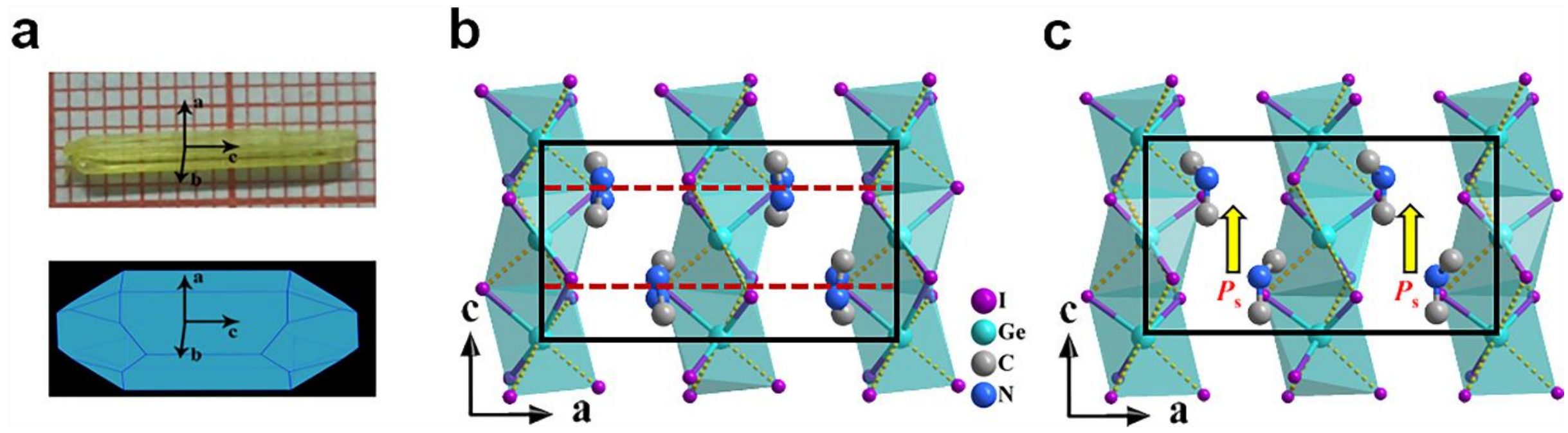

**Fig. 1 | Structural information of $DMAGeI_3$. a**, The image and simulated crystal growth morphology of single crystal. **b-c**, Packing view of the unit cell in HTP and LTP along the *b*-axis. Hydrogen ions are omitted for clarity. The red dashed lines indicate the 2-fold crystallographic rotation axes and the polarization direction are indicated by yellow arrows.

The structural analysis discloses that $DMAGeI_3$ crystallizes in a centrosymmetric space group *Pnan* in HTP, while in LTP it transforms into a polar space group $Pna2_1$. Figure 1b, c show that the two phases are similar except for the DMA orientation and $GeI_3$ distortion. $DMAGeI_3$ adopts the hexagonal perovskite structure containing 1D inorganic framework, where the $GeI_3$ units are linked via the face-sharing $GeI_6$ octahedra. The cavities between $GeI_3$ columns are filled by organic DMA cations, which are loosely connected with the inorganic framework via N-H···I and C-H···I hydrogen bonding interactions. In HTP, the organic DMA cations locate at the 2-fold crystallographic rotation axes of lattice and exhibit orientational disorder toward two equivalent directions, making their dipole moments cancel each other (Fig. 1b). In LTP, the DMA cations become ordered with an aligned manner along the *c*-axis (Fig. 1c and Supplementary Fig. 5), and also their positions move significantly (0.38 Å) along the *c*-axis direction compared with those in HTP (Supplementary Fig. 6), both of which induce a net dipole moment in this direction (Fig. 1c). Concomitantly, the Ge-I bond lengths and I-Ge-I angles in LTP also changed obviously (Supplementary Fig. 7 & Supplementary Table 2), although the $GeI_6$ octahedra are already distorted in HTP. The value of the octahedron distortion parameter Δ and angle variance $\sigma_{oct}$ rise from 0.0096/60.6 (HTP) to 0.0121/69.3 (LTP) respectively, which reflect a more distorted state of $GeI_6$ octahedra in LTP (Supplementary Table 3).

These structural characteristics, including the orientation ordering of DMA cation, the structural distortion of $GeI_3$ framework, and the relative moving between these two parts, will contribute to a ferroelectric polarization along the *c*-axis, which will be further analyzed later.

**Characterization of ferroelectricity**

According to the above structural analysis, the symmetry breaking occurs with the change of space group from *Pnan* to $Pna2_1$, and the symmetry elements halved from eight ($E$, $C_2$, $2C_2$, $i$, $\sigma_h$, and $2\sigma_v$) to four ($E$, $C_2$, and $2\sigma_v$), which means the $DMAGeI_3$ is a *mmm*F*mm*2-type ferroelectric with two equivalent polarization directions[42].

Second harmonic generation (SHG) experiment on powder sample was used to characterize its paraelectric-ferroelectric phase transition (Supplementary Note 2). As shown in Fig. 2a, $DMAGeI_3$ shows a strong SHG response at room temperature, which confirms its non-centrosymmetric LTP. With increasing temperature, the SHG signal intensity gradually decreases to zero above $T_C$, indicating that the HTP structure is transformed into a centrosymmetric one. Moreover, the SHG measurement on a single crystal surface of $DMAGeI_3$ can further reveal the direction vector of ferroelectric polarization (Supplementary Note 2 & Supplementary Fig. 9)[43,44]. As shown in Fig. 2b, the angle-dependent SHG intensity shows a clear bipolar behavior, namely its intensity reaches the maximum (minimum) when the polarization direction of incident light is parallel (perpendicular) to the spontaneous polarization direction (i.e., $c$-axis) of $DMAGeI_3$ crystal. It is worth to mention that the SHG strength of $DMAGeI_3$ (of power sample) is more than ten times that of KDP (Fig. 2a), comparable to that of previous reported hybrid halide antiperovskite $Cs_3Cl(HC_3N_3S_3)$ (11.4 × KDP)[45], and better than many other reported ferroelectrics (Fig. 2c and Supplementary Table 4). Thus, a direct application of $DMAGeI_3$ crystal is the nonlinear optical converter, which can omit bright green light via photoluminescence, as demonstrated in Supplementary Fig. 10.

Paraelectric-ferroelectric phase transition is generally accompanied by an obvious dielectric anomaly. The temperature-dependent dielectric real part ($\varepsilon'$) of $DMAGeI_3$ was measured on single crystals along three crystallographic axes at various frequencies. As shown in Figs. 2d and 2e, the dielectric constant along the $c$-axis ($\varepsilon'_c$) is much larger than those along the other two axes, especially in the region around $T_C$. This dielectric anisotropy can be ascribed to this $mmm$F$mm2$-type phase transition that only allows spontaneous polarization along the $c$-axis. It is noteworthy that the value of $\varepsilon'_c$ reveals an obvious frequency-dependence. Specifically, as the frequency decreases from 1 MHz to 1 kHz, the peak value of $\varepsilon'_c$ at $T_C$ increases from 4679 to 102988 which is about 3269 times of its dielectric constant (~31.5) at room temperature (Fig. 2e). To our knowledge, this is the largest dielectric response in the reported hybrid perovskite materials to date (Fig. 2f & Supplementary Table 5). Additionally, according to the Curie-Weiss law, the $C_{para}$ and $C_{ferro}$ are calculated to be 9664 and 17469 K at 1 kHz, respectively. The ratio of $C_{ferro}$ /$C_{para}$ is 1.8 (smaller than 4), which discloses the characteristics of second-order ferroelectric phase transition (Supplementary Fig. 11).

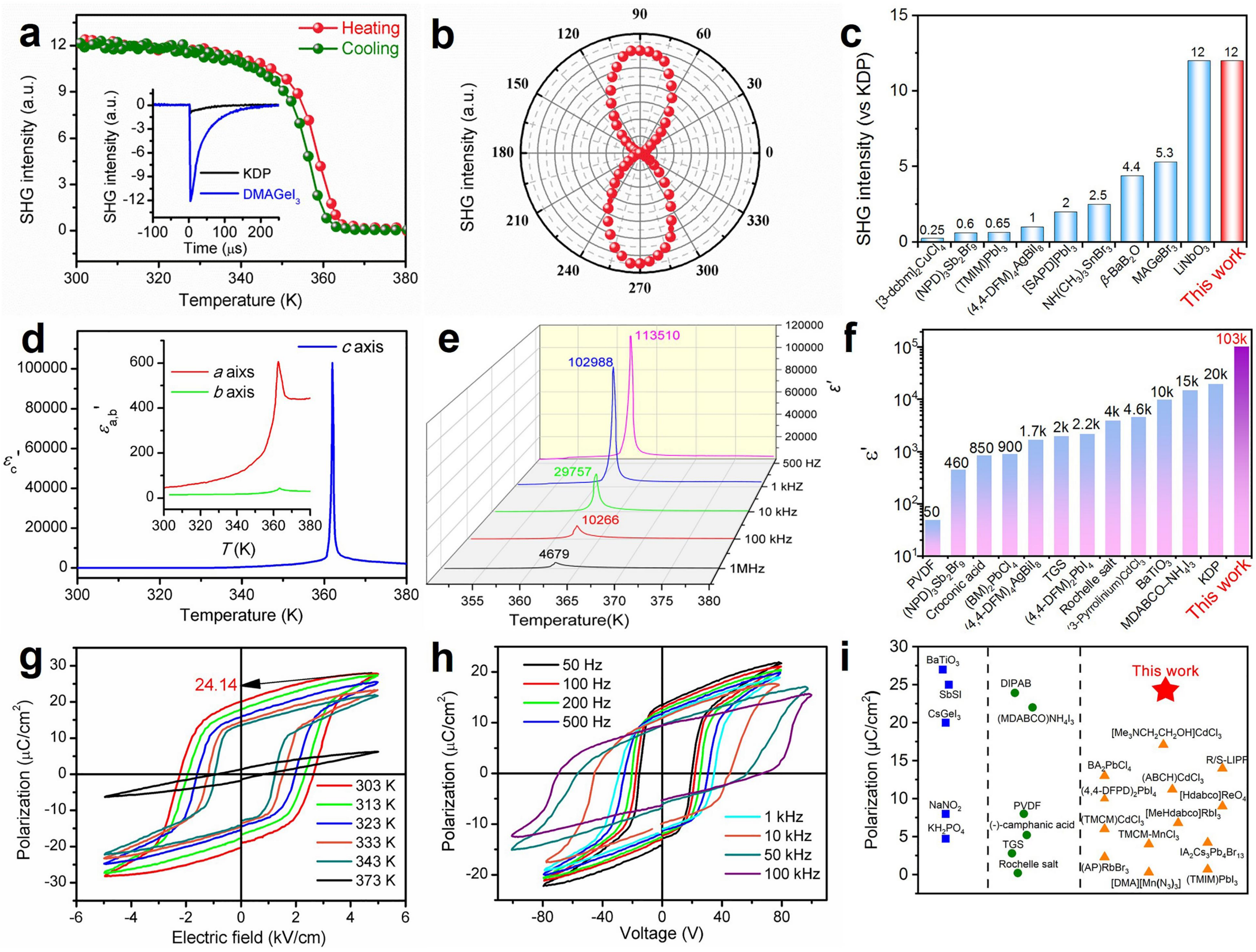

**Fig. 2 | Ferroelectric properties of DMAGeI$_3$. a**, The SHG intensity of DMAGeI$_3$ as a function of temperature on powder samples. Inset: the comparison of SHG signals between DMAGeI$_3$ and KDP powder samples (in the same range of particle diameter 300-450 μm) at room temperature. More comparisons with various particle sizes can be found in Supplementary Fig. 8, which leads the same conclusion. **b**, The SHG anisotropic polar plot of DMAGeI$_3$ crystal, where the $\varphi$ is the angle between the *b*-axis and the polarization of incident light. **c**, Comparison of SHG intensity between DMAGeI$_3$ and other ferroelectrics. **d**, The temperature-dependent dielectric real part ($\varepsilon'$) of DMAGeI$_3$ measured on single crystals along the *a*, *b*, and *c*-axis at 1 kHz. **e**, The temperature-dependent dielectric real part ($\varepsilon'$) of DMAGeI$_3$ measured on single crystals along the *c*-axis at various frequencies. **f**, Comparison of the maximum value of $\varepsilon'$ of DMAGeI$_3$ and other ferroelectrics. **g-h**, The typical polarization-electric field (*P-E*) hysteresis loops measured along the *c*-aixs at various temperatures and at different frequencies. **i**, Comparison of polarization values between DMAGeI$_3$ and other ferroelectrics.

The polarization-electric field ($P$-$E$) hysteresis loop is the most direct evidence of ferroelectricity. Fig. 2g shows the $P$-$E$ loops at various temperatures. Above $T_C$ (e.g. 373 K), the polarization response to the applied field presents almost linear behavior with a very narrow loop, while the well-defined $P$-$E$ hysteresis loops were observed in the range of 343-303 K (below $T_C$), as expected for the paraelectricity and ferroelectricity of the HTP and LTP. As the temperature decreases in the LTP, both the saturation polarization ($P_s$) and remnant polarization ($P_r$) increase gradually, and reach up to 24.14 μC/cm$^2$ and 20.26 μC/cm$^2$ at 303 K, respectively. Remarkably, the $P_s$ of $DMAGeI_3$ is larger than those of most reported famous organic and OIHP ferroelectrics such as $[Me_3NCH_2CH_2OH]CdCl_3$ (17.1 μC/cm$^2$)[28], $[MeHdabco]RbI_3$ (6.8 μC/cm$^2$)[29], $[Me_3NCH_2Cl]MnCl_3$ (4.0 μC/cm$^2$)[38], and very close to that of $BaTiO_3$ (Fig. 2i & Supplementary Table 6).

Another notable feature of $DMAGeI_3$ ferroelectricity is its small coercive field ($E_c$) values (0.8-2.2 kV/cm), which are one order of magnitude lower than that of the recently reported all-inorganic perovskite $CsGeI_3$ (~40 kV/cm)[41], and two orders of magnitude lower than those of PVDF (~500 kV/cm) and its copolymers[39]. More comparisons with other hybrid ferroelectrics can be found in Supplementary Table 7. Such a small $E_c$ is advantageous for energy-saving operations in devices. Furthermore, the polarization switching speed is also a key property for ferroelectric applications, thus the frequency dependence of $P$-$E$ hysteresis loops is investigated, as shown in Fig. 2h. With rising frequency, the values of $P_s$ and $P_r$ slightly decrease, while the coercive voltage increases. Its ferroelectric polarization can be switched at a frequency up to 100 kHz, which shows excellent high-frequency performance for macroscopic crystals with large scale domains. Given the large $P_s$ value in combination with the small $E_c$ and the high-frequency performance, $DMAGeI_3$ represents promising hybrid ferroelectric materials for device applications. Moreover, the piezoelectric response has also been measured by using the quasi-static method. At room temperature, the $d_{33}$ of the $DMAGeI_3$ crystal reaches 3 pC/N, along the polar axis [001] under a testing frequency of 110 Hz (Supplementary Fig. 13).

For ferroelectrics, the microscale domain structure is another essential property of polarity, which can be characterized by using the piezoresponse force microscopy (PFM) technology[46].

By adjusting the lateral (in-plane) or vertical (out-of-plane) mode in PFM, the relative strength of piezoelectric response and polarization directions from different components can be harvested. Figs. 3a and 3b display a schematic diagram of PFM in-plane testing on $DMAGeI_3$ crystal along the *c*-axis (polarization direction), and the obtained topography of the crystal surface is shown in Fig. 3c. In the lateral mode, the observed obvious stripy domain walls parallel to the *c*-axis eliminate the interference of topography and confirm the existence of ferroelectric polarization (Fig. 3d). Combined with amplitude, the two sides of the domain wall show the response to different polarization directions, displaying obvious phase difference (Fig. 3e). In contrast, the piezoelectric signals do not give a component in the vertical direction (Supplementary Fig. 14), verifying the results of crystal structure determination (*mmm*F*mm*2-type ferroelectric).

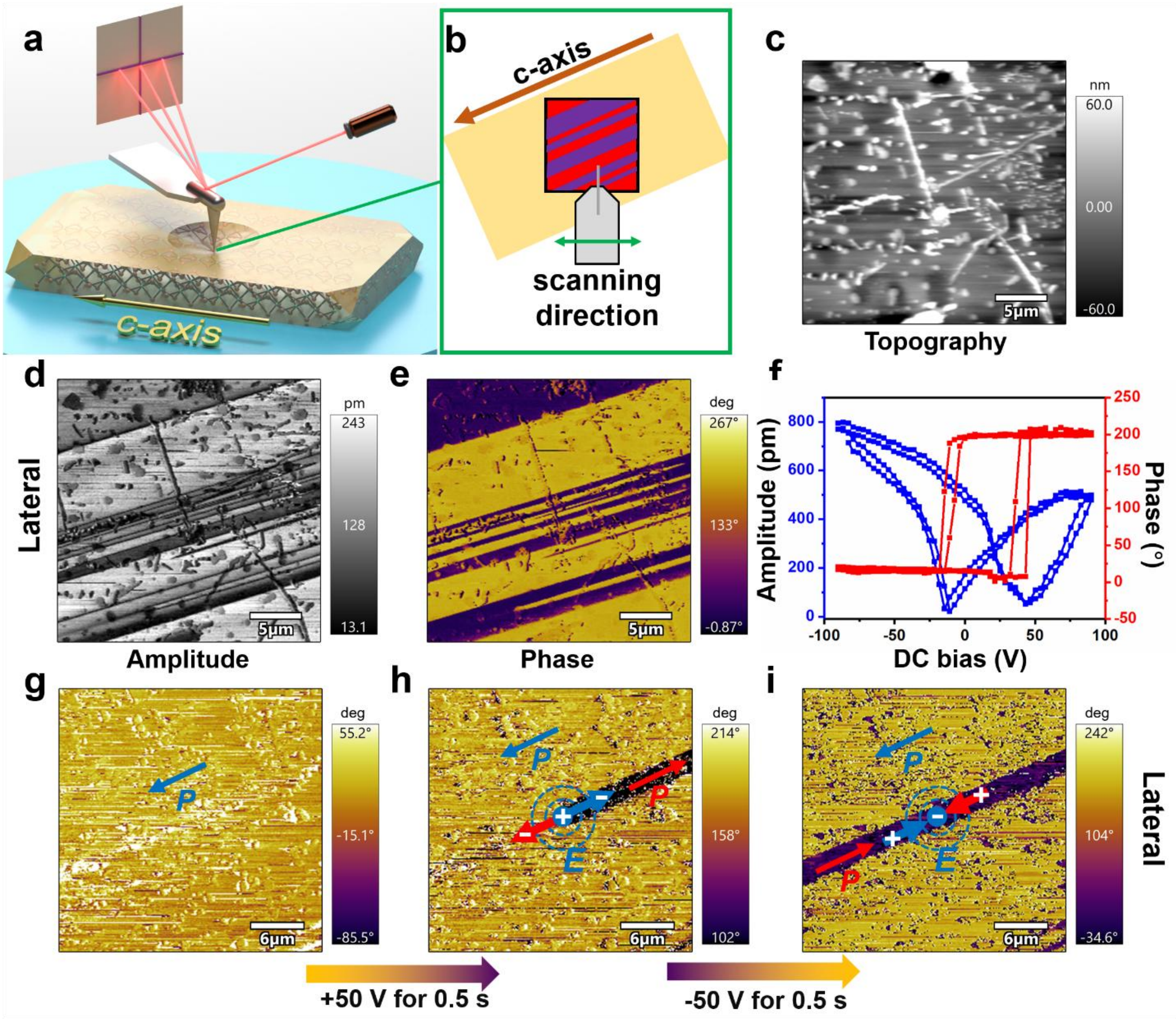

**Fig. 3 | Ferroelectric domain structures and polarization switching of $DMAGeI_3$. a-b**, Three-dimensional and overhead schematic diagram of lateral mode of PFM for $DMAGeI_3$. **c-e**, Topography, lateral amplitude, and phase images. **f**, Lateral PFM amplitude (blue) and phase (red) signals as functions of the tip voltage, showing local PFM hysteresis loops. **g**, Initial phase image with 30 × 30 $\mu m^2$. **h**, Phase image after applying positive bias of +50 V for 0.5 s in the center of the area. **i**, Succeeding phase image after applying negative bias of -50 V for 0.5 s at the same location.

To further illustrate the ferroelectricity of $DMAGeI_3$, the ferroelectric domain is switched using PFM. As shown in Fig. 3f, the butterfly-shaped amplitude (blue) and phase hysteresis loops (red) provide evidence for ferroelectric switching. A single domain state region with 30× 30 $\mu m^2$ is selected to demonstrate the domain inversion of $DMAGeI_3$ (Fig. 3g). After applying positive bias of +50 V for 0.5 s, in the sectional area parallel to the *c*-axis on the right side of the point where the bias was applied, the polarization direction is switched (Fig. 3h). On the contrary, when giving a bias of -50 V for 0.5 s, the domain in the left area is reversed, resulting in the appearance of the stripy domain (Fig. 3i). Subsequently, a higher bias is applied to the center point, and the left area turns to the same phase, which is obviously different from the reverse polarization direction on the right (Supplementary Fig. 15). Our PFM measurement results strongly prove that $DMAGeI_3$ possesses stable and switchable polarization.

## DISCUSSION

To further understand the prominent ferroelectricity of $DMAGeI_3$, we performed density-functional-theory (DFT) calculations to reveal underlying mechanisms. Our DFT calculation leads to lattice constants close to the experimental LTP one (Supplementary Fig. 17). And its electronic structure is shown in Supplementary Fig. 18, suggesting a good insulator with a band gap 2.79 eV. The valance band maximum is contributed by Ge and I. The theoretical value of polarization is estimated as 29.09 $\mu C/cm^2$ at the ground state (i.e., $T$ = 0 K), which agrees well with (and slightly higher than) the experimental value at room temperature.

Then the ferroelectric origin of $DMAGeI_3$ is analyzed. First, the $4s^2$ lone pair of $Ge^{2+}$ ion is a strong driving force of ferroelectricity, as appeared in the all-inorganic 3D perovskite $CsGeI_3$[41]. As illustrated in Fig. 4a, the $4s^2$ lone pairs form a zigzag pattern along the inorganic chain, leading to an uncompensated dipole along the *c*-axis in the LTP. In contrast, in the HTP structure, the local dipoles from $4s^2$ lone pairs are fully compensated, as compared in Fig. 4b. Second, besides the contribution from $Ge^{2+}$'s $4s^2$ lone pair, the effect of DMA cation is also non-negligible. Both the bodily movement of this +1 charged group and stereo orientation of this asymmetric structure will generate a local dipole. A simple estimation of their individual contributions are 16.18 μC/cm$^2$ (from the DMA cations) and 13.62 μC/cm$^2$ (from the $GeI_3$ framework) respectively (see Supplementary Note 3 and Supplementary Fig. 19). In other words, the large polarization of $DMAGeI_3$ is due to the synergistic effect of inorganic and organic parts, with almost half-half weights.

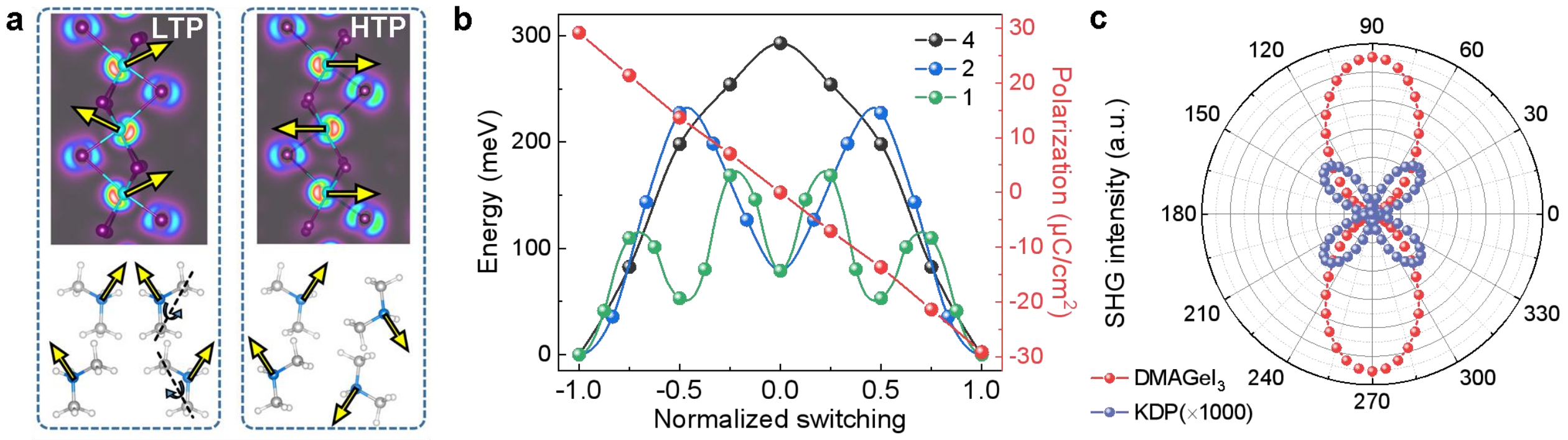

**Fig. 4 | Ferroelectricity revealed by *ab initio* calculations. a**, The schematic of polarization of the $GeI_3$ framework (upper) and DMA cations (lower). The yellow arrows correspond to the local dipoles' directions. The lone pairs of Ge's $4s^2$ electron clouds can be clearly visualized as a driving force of dipoles. And the dipoles of DMA cations can be flipped by rotation. **b**, The calculated energy barriers and evolution of polarization of ferroelectric switching. The barrier depends on the switching process, namely how many DMA cations (1, 2, or 4) rotate synchronously in a unit cell. The one-by-one rotation of DMA rotation leads to a lower barrier. Note that the theoretical barrier only defines the upper limit, while the real barrier can be even lower, especially at room temperature. **c**, The comparison of calculated SHG plots of $DMAGeI_3$ and KDP crystals (both on the *bc* plane). The 0° (90°) is along the *b*-axis (*c*-axis).

The ferroelectric switching process of $DMAGeI_3$ can be also simulated by rotating the DMA cations, as depicted in Fig. 4a. If the four DMA cations in one unit cell rotate synchronously, the switching barrier energy is 292.9 meV, which is not as low as expected. However, this barrier can be significantly reduced by rotating the DMA cations one by one, as compared in Fig. 4b, where the first and second barriers of the rotating steps are 109.6 meV and 115.6 meV, respectively. For comparison, the DFT switching barrier for a $BaTiO_3$ unit cell is ~8.3 meV and its experimental coercive field $E_c$ is <10 kV/cm[12]. But it should be noted that the unit-cell volume of $DMAGeI_3$ is ~16.5 times of $BaTiO_3$. Thus, the two one-by-one barriers are only ~6.6 meV and ~7.0 meV for $DMAGeI_3$ when normalized to the volume of $BaTiO_3$. Compared with $BaTiO_3$, the proximate $P_s$ and lower barrier of $DMAGeI_3$ suggest an even smaller $E_c$, as observed in our experiments.

Furthermore, its prominent SHG signal can be also well reproduced theoretically. The nonlinear optical susceptibility tensors ($d_{ij}$'s) of LTP of $DMAGeI_3$ are calculated, which determine the intensity and anisotropy of SHG signals. For $DMAGeI_3$ with space group $Pna2_1$, its SHG intensity on the *bc* crystalline plane can be expressed as[47]: $I_{DMAGeI_3} \propto (2d_{24}\sin 2\varphi)^2 + (2d_{32}\cos^2\varphi + 2d_{33}\sin^2\varphi)^2$, where $\varphi$ is the angle between the polarization direction of incident light and *b*-axis (more details can be found in Supplementary Note 4). As shown in Fig. 4c, the calculated SHG plot of $DMAGeI_3$ shows a bipolar behavior, and the strongest signal appears when the polarization direction of incident light is along the *c*-axis of crystal, consistent with the experimental observation.

For comparison, the $d_{ij}$'s of KDP room temperature phase (space group $I\bar{4}2d$) is also calculated, which agrees well with the experimental values (Supplementary Fig. 20). As expected, these values are systematically smaller than those of $DMAGeI_3$. Thus the SHG signal of KDP is much smaller than that of $DMAGeI_3$, since generally the SHG response is proportional to the square of $d_{ij}$'s, as compared in Fig. 4c.

Our study unambiguously demonstrated the excellent comprehensive ferroelectric properties of $DMAGeI_3$ with quasi-one-dimensional structure, which is a breakthrough of organic-inorganic hybrid ferroelectrics as well as low-dimensional ferroelectrics. Its remarkable ferroelectricity originates from the synergistic overlap of two robust mechanisms: ordering of polar cations and soft mode of $4s^2$ lone pairs, which doubles its net polarization and keeps its coercivity at a low level. This successful example provides a reliable and simple route to pursue more superior ferroelectrics or even multiferroics in organic-inorganic hybrid systems, to finally reach the aim of commercial applications. The opportunities are vast for emerging hybrid germanium halide perovskites, and the ones equipped with multipolar characteristics and outstanding piezoelectric performance may be available in the near future.

## Methods

**DSC, SHG, and XRD measurements.** Differential scanning calorimetry (DSC) measurements were carried out by using a NETZSCH DSC 200F3 instrument under the nitrogen atmosphere, where the dry powder of $DMAGeI_3$ (28.9 mg) were heated and cooled with a rate of 10 K/min in the temperature ranges of 293-390 K. SHG switching experiments were measured with powder samples by pulsed Nd: YAG (1064 nm, Vibrant 355 II, OPOTEK), and the temperature varies from 300 K to 380 K. The powdered $KH_2PO_4$ (KDP) was used as the reference. The SHG phase matching tests were performed with a particle size range of 62-375 μm at room temperature. Furthermore, the SHG intensity anisotropy was investigated by using Witec alpha 300 on single-crystal surfaces. Variable-temperature single-crystal X-ray diffraction data were collected on a Rigaku VarimaxTM DW diffractometer with Mo Kα radiation (λ = 0.71073 Å). Data processing with empirical absorption correction was conducted by using the CrysAlisPro 1.171.40.14e (Rigaku OD, 2018). The crystal structures were confirmed by direct methods and refined by full-matrix least-squares methods based on $F^2$ through the OLEX2 and SHELXTL (version 2018) software package. All non-hydrogen atoms were refined anisotropically and all hydrogen atoms were generated geometrically in suitable positions.

**Dielectric and ferroelectric measurements.** For the dielectric measurements, the sample of $DMAGeI_3$ was made with single-crystals cut perpendicular to the *a*-, *b*-, and *c*-axis respectively. Silver conduction paste deposited on the plate surfaces was used as the electrodes. The temperature-dependent dielectric constants were performed on the Tonghui TH2828A instrument under the frequency range from 1 kHz to 1 MHz with an applied electric field of 1 V. The single crystal electrodes of $DMAGeI_3$ were prepared for the *P-E* hysteresis loop measurements by Sawyer-Tower method. The ferroelectric polarization imaging and local switching studies on the bulk crystal surface were carried out by using a resonant-enhanced PFM (MFP-3D, Asylum Research).

**PXRD and TGA measurements.** Powder XRD data were measured using a Rigaku D/MAX 2000 PC X-ray diffraction system with Cu Kα radiation in the 2θ range of 5°–50° with a step size of 0.02°. The thermogravimetric analysis (TGA) were performed on the NETZSCH TG 209F3 instrument in the range of 298–1000 K with a heating rate of 20 K/min.

**DFT Calculations.** The electronic structure calculations were performed based on projector augmented wave pseudopotentials, as implemented in Vienna *ab initio* Simulation Package (VASP)[48]. Plane-wave cutoff energy of 500 eV and 3×4×4 *k*-point meshes were used. The lattice constants and the atomic positions were fully relaxed with the convergence criteria of $10^{-5}$ eV and 0.01 V/Å for energy and force, respectively. The exchange-correlation potential was approximated in the form of Perdew-Burke-Ernzerhof (PBE) functional and the van der Waals (vdW) interaction was considered by adding D3 Grimme correction[49,50]. Other choices of functional and vdW correction were also tested (Supplementary Fig. 17). The climbing nudged elastic band method was used to determine the paths and energy barrier of ferroelectric switching[52]. Ferroelectric polarization was calculated by using the Berry phase method[53]. The second-harmonic nonlinear optical susceptibility tensors for SHG were calculated using the exciting package[53].

**Data Availability**

The experimental cif files can be found in CCDC ($DMAGeI_3$: 2227301 (223 K), 2227302 (253 K), 2213657 (293 K), 2220591 (373 K); $DMAGeCl_3$: 2213654 (293 K); $DMAGeBr_3$: 2213655 (293 K)). Source data is provided with this paper.

**Acknowledgments**

This work is supported by National Natural Science Foundation of China (Grant Nos. 21991141, 92056112, 12274069, 11834002).

**Author Contributions**

D.-W.F.,Y.Z., S.D. and Z.-X.Z. conceived the project. D.W.F. and Y.Z. designed the experiments. S.D. proposed the theoretical mechanisms. K.D. prepared the samples and performed the DSC, single crystal measurement and analysis. Z.-X.Z. contributed to dielectric and *P-E* loops measurements and analysis. H.Y. performed the DFT calculations guided by S.D.. C.S. contributed to the analysis of PFM. Y.A.X., D.W., G.D. and Y.M.Y. contributed to PFM and SHG measurements. Z.-X.Z., D.W.F., S.D. and Y.Z. wrote the manuscript, with inputs from all other authors.

**Competing interests:** The authors declare no competing interests.

# Supporting Information

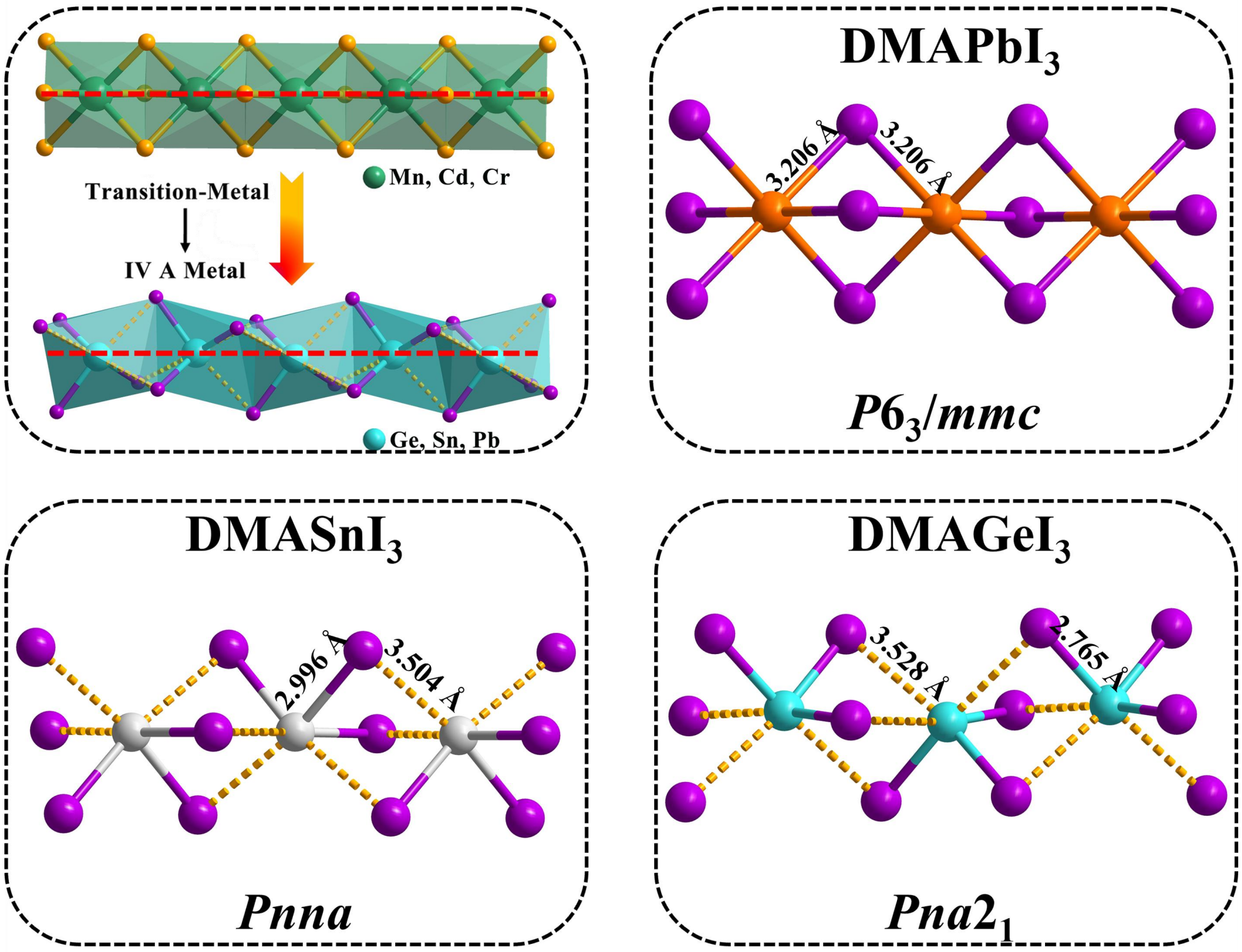

**Supplementary Fig. 1 | | The Strategy for designing Ge-based molecular ferroelectric materials.** The structures of $DMAPbI_3$[1], $DMASnI_3$[2], and $DMAGeI_3$ are obtained at 293 K.

**Supplementary Note 1 | More details of synthesis and crystal growth.**

*N, N*-Dimethylformamide (DMF, 97%, Macklin), $GeO_2$ (99%, Macklin), HBr (40%, Macklin), HCl (36%, Adamas), $H_3PO_2$ (50 wt.% in $H_2O$, Macklin) and HI (50.0%, Macklin). All chemicals were commercially available and were used without further purification.

$DMAGeCl_3$ was prepared by dissolving 2 mmol of $GeO_2$ in a mixed solution of 10 mL HCl, 5 mL $H_3PO_2$ and 5 mL DMF. First, 2 mmol (0.209 g) $GeO_2$ was slowly added into the mixed solution of 10 mL HCl and 5 mL $H_3PO_2$. A clear solution was obtained after strong stirring for 2 h at 373 K. Subsequently, 5 mL DMF was added to the solution which continued to stir for 1 h. After the reaction mixture was cooled to room temperature and filtered, the colorless solution was kept at room temperature for slow evaporation. The colorless block crystals were obtained about a few days later. Following a similar synthetic procedure, the colorless block crystals $DMAGeBr_3$ could also be obtained (Supplementary Fig. 16). The isostructural compounds $DMAGeCl_3$ and $DMAGeBr_3$ crystallize in the orthorhombic system of space group *Pbca* at room temperature. More detailed structural information is summarized in Supplementary Tables 8-9.

However, the $DMAGeI_3$ crystals could not be obtained under the above experimental conditions. We successfully obtained $DMAGeI_3$ single crystals through the hydrothermal method. The $GeO_2$ (2 mmol) was dissolved in a mixed solution of 3 mL HI, 3 mL $H_3PO_2$, and 3 mL DMF. The prepared suspension was then transferred into 25 mL Teflon-lined stainless-steel autoclaves and maintained at 363 K for 6 h. Then, the saturated solution was slowly cooled with a temperature rate of 5 K/day. Finally, the yellow rod-shaped crystals were successfully obtained after cooling the autoclave to room temperature. The $DMAGeI_3$ crystallizes in a polar space group $Pna2_1$ at room temperature. More detailed structural information is summarized in Supplementary Tables 1-2.

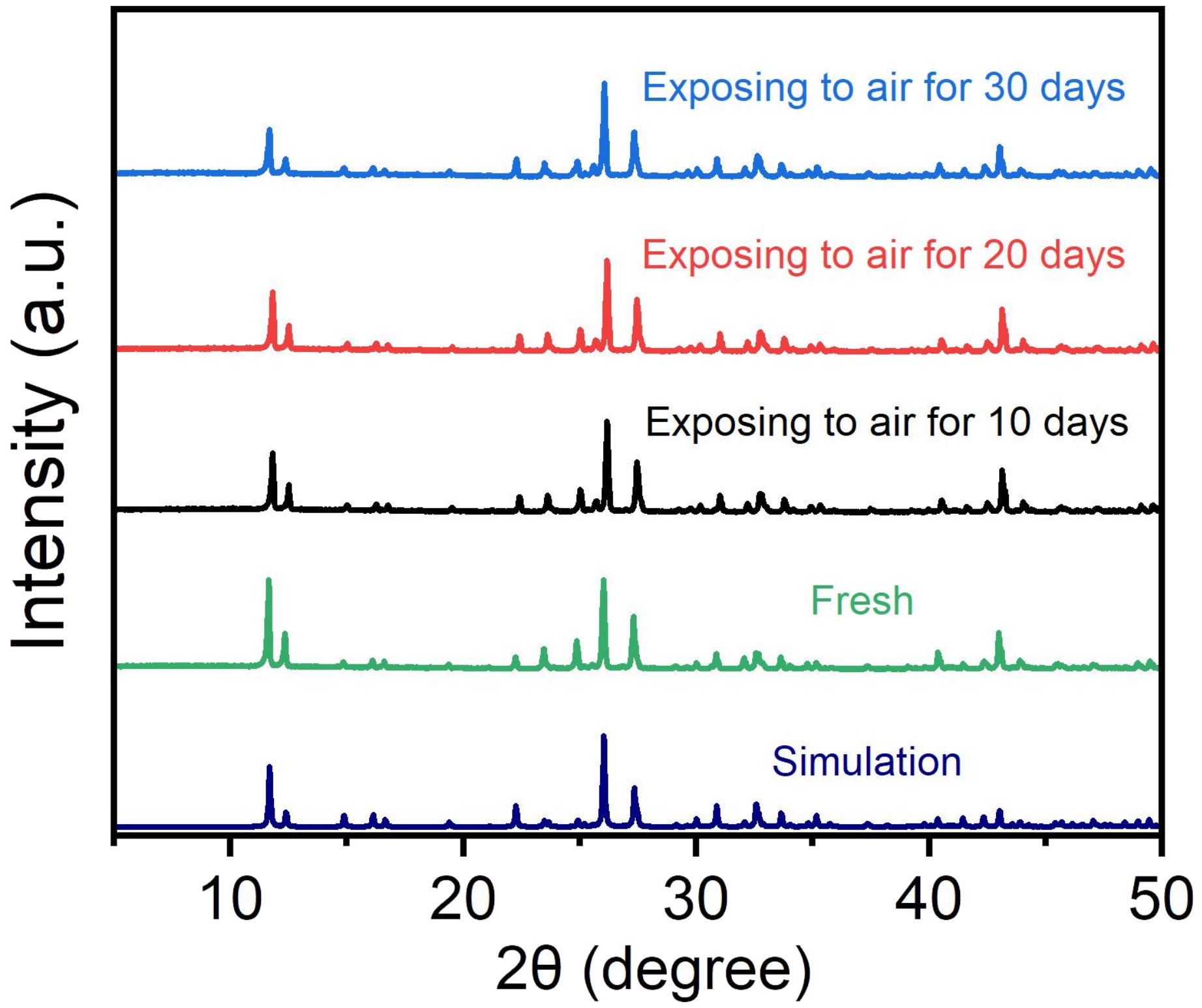

**Supplementary Fig. 2 | Powder X-ray diffraction patterns of $DMAGeI_3$ in different test conditions.**

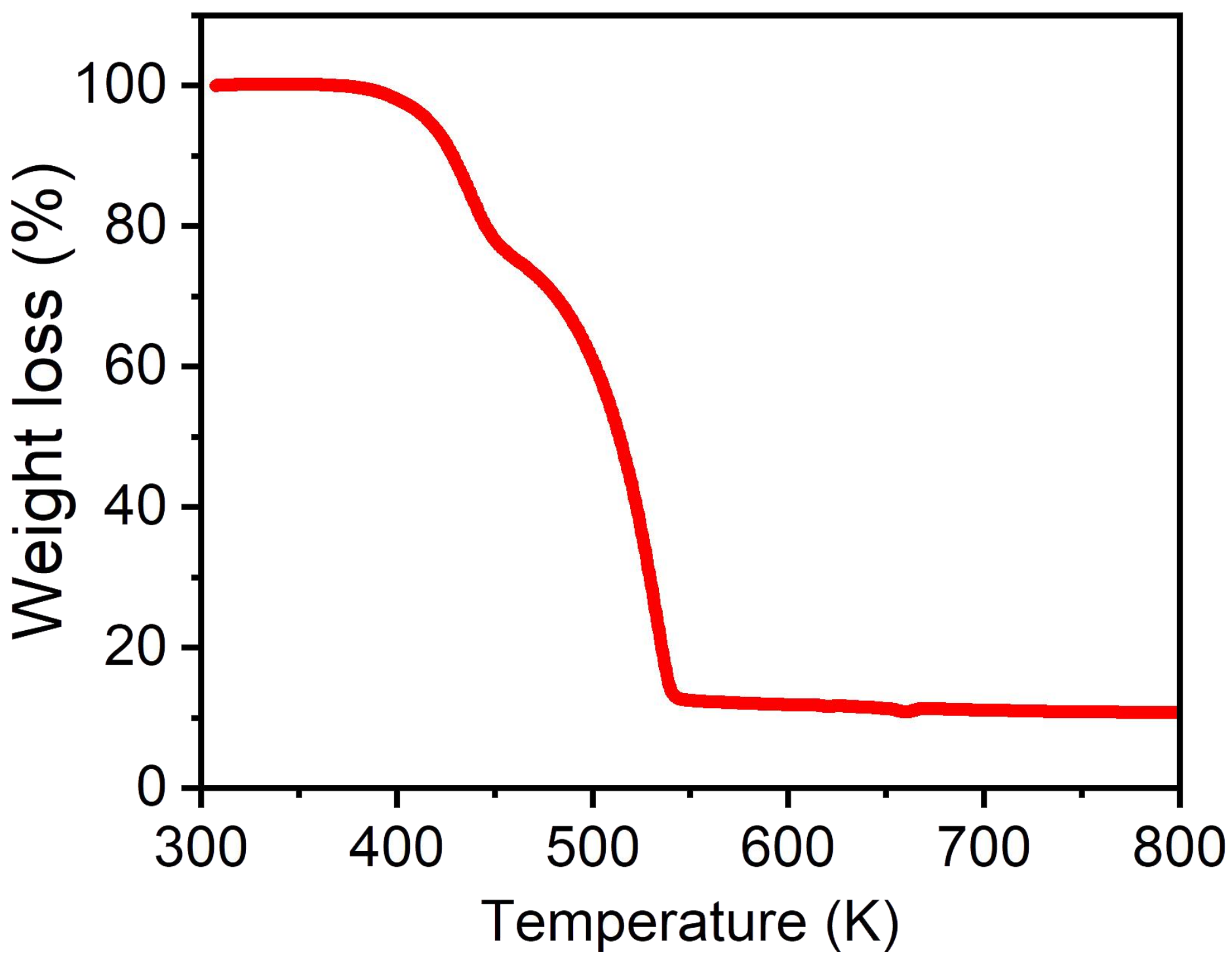

**Supplementary Fig. 3 | TGA curve of $DMAGeI_3$.**

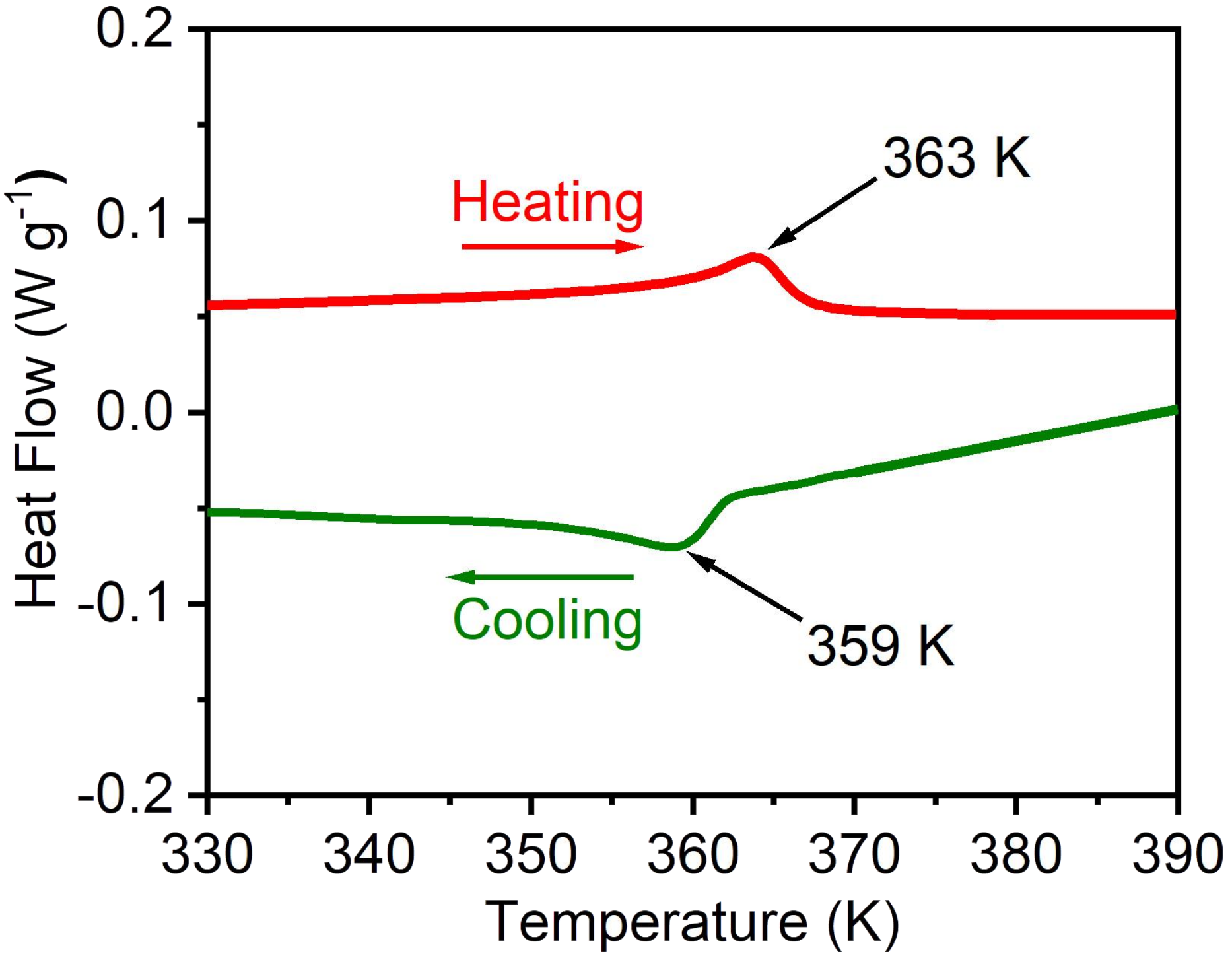

**Supplementary Fig. 4 | Differential Scanning Calorimeter (DSC) curves of $DMAGeI_3$.**

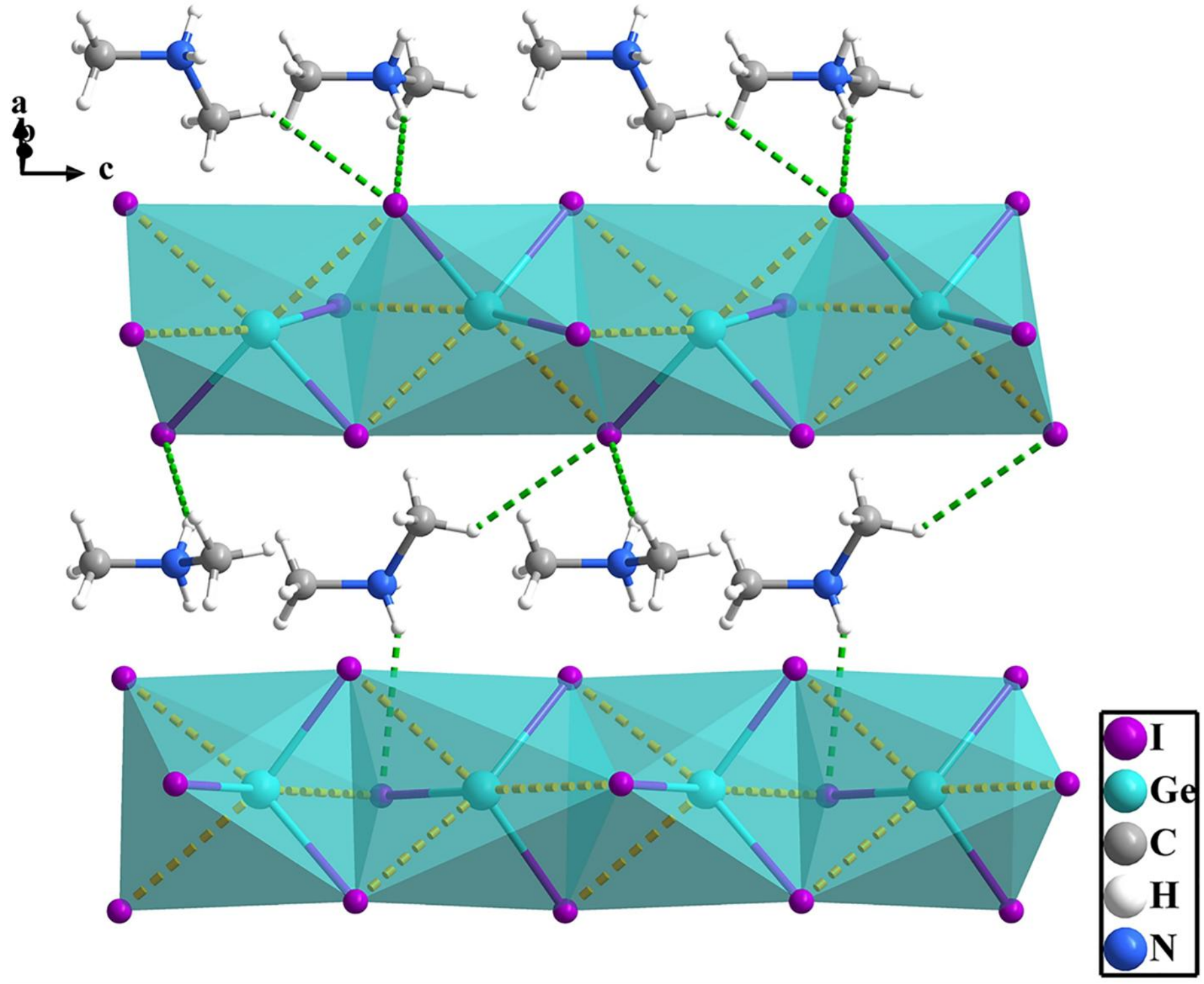

**Supplementary Fig. 5 | The crystal structural analysis of $DMAGeI_3$.** The N-H···I hydrogen-bonding interactions between organic cations and inorganic perovskite frameworks of $DMAGeI_3$ at 293 K.

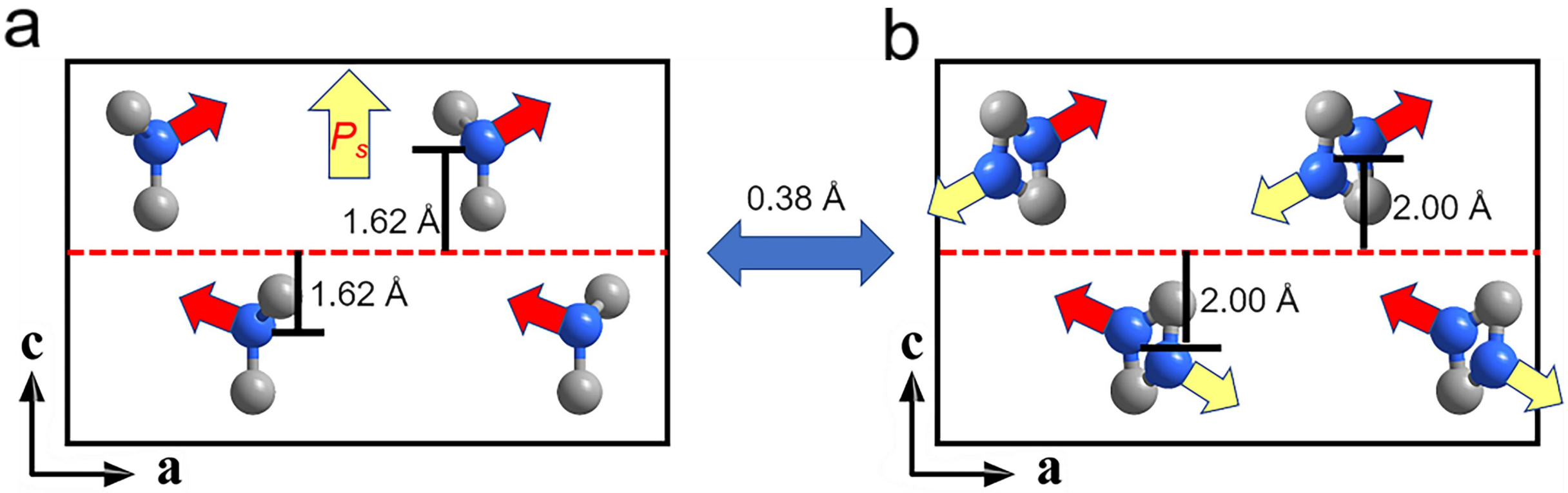

**Supplementary Fig. 6 | Comparison of DMA cations. a-b**, The relative displacement of DMA cations along the *c*-axis at 293 K, 373 K. The red dashed lines indicate the horizontal center position in the unit cell and the local dipoles are indicated by arrows.

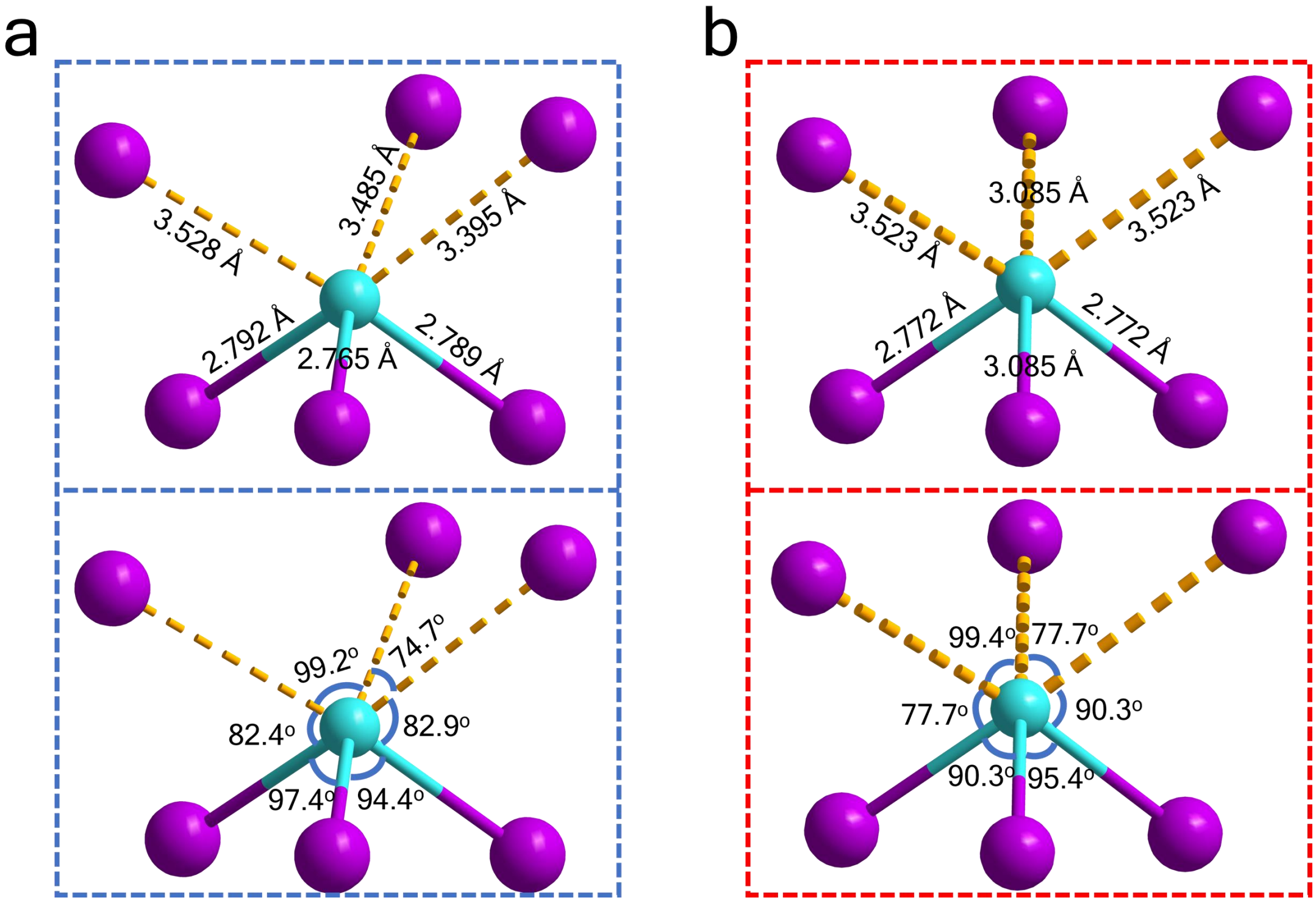

**Supplementary Fig. 7 | Comparison of bond lengths and bond angles of $GeI_6$ octahedron.** **a-b**, 293 K and 373 K, respectively.

**Supplementary Table 1 | Crystal data and structure refinement details for compound DMAGeI$_3$.**

| DMAGeI$_3$ | 223 K | 253 K | 293K | 373 K |
|---|---|---|---|---|
| Empirical formula | $C_2H_8NGeI_3$ | $C_2H_8NGeI_3$ | $C_2H_8NGeI_3$ | $C_2H_8NGeI_3$ |
| Formula weight | 499.40 | 499.40 | 499.40 | 499.40 |
| Crystal system | orthorhombic | orthorhombic | orthorhombic | orthorhombic |
| Space group | $Pna2_1$ | $Pna2_1$ | $Pna2_1$ | *Pnan* |
| a/Å | 14.1481(4) | 14.1793(7) | 14.2829(10) | 14.5493(6) |
| b/Å | 8.8837(3) | 8.9119(4) | 8.9305(8) | 8.9096(4) |
| c/Å | 7.9434(3) | 7.9572(4) | 7.9828(6) | 8.0080(3) |
| $\alpha$/° | 90 | 90 | 90 | 90 |
| $\beta$/° | 90 | 90 | 90 | 90 |
| $\gamma$/° | 90 | 90 | 90 | 90 |
| Volume/Å$^3$ | 998.39(6) | 1005.51(8) | 1018.23(14) | 1038.07(7) |
| Z | 4 | 4 | 4 | 4 |
| $F$ (000) | 872.0 | 872.0 | 868.0 | 840.0 |
| Radiation | Mo Kα | Mo Kα | Mo Kα | Mo Kα |
| absorption correction | emiempirical | emiempirical | emiempirical | emiempirical |
| GOF | 1.070 | 1.100 | 1.047 | 1.101 |
| $R_1$ | 0.0774 | 0.0789 | 0.0822 | 0.0440 |
| $wR_2$ | 0.2024 | 0.2147 | 0.1939 | 0.1143 |

**Supplementary Table 2 | Selected bond lengths [Å] and angles [°] for compounds DMAGeI$_3$ at 293 and 373 K, respectively.**

| 293 K | | 373 K | |
|---|---|---|---|
| I001—Ge04 | 2.792 (3) | I001—Ge03 | 2.7722 (10) |
| I002—Ge04 | 2.765 (3) | I002—Ge03 | 3.0852 (7) |
| I003—Ge04 | 2.789 (3) | C2—N1$^{i}$ | 1.35 (2) |
| C005—N1 | 1.52 (4) | C2—N1 | 1.49 (3) |
| C006—N1 | 1.45 (4) | | |
| | | I001$^{ii}$—Ge03—I001 | 95.46 (4) |
| I002—Ge04—I001 | 97.42 (10) | I001$^{ii}$—Ge03—I002 | 94.77 (2) |
| I002—Ge04—I003 | 94.51 (10) | N1$^{i}$—C2—N1 | 61.7 (15) |
| I003—Ge04—I001 | 94.48 (9) | C2$^{i}$—N1—C2 | 114.7 (15) |
| C006—N1—C005 | 114 (3) | I001—Ge03—I002 | 90.39 (2) |
| | | N1$^{i}$—C2—N1—C2$^{i}$ | −22 (2) |

**Supplementary Table 3 | The Ge-I bond length (Å) and I-Ge-I bond angle (°) of the GeI$_6$ octahedron at 293 and 373 K, respectively.** The octahedron distortion parameter $\Delta = \frac{1}{6}\sum_{i=1}^{6}\left(\frac{d_i-d}{d}\right)^2$, *di*, *d* are the individual bond length and the average length of the Ge-I in the

[$GeI_6$] octahedron. The octahedral angle variance $\sigma^2_{\text{oct}} = \frac{1}{11}\sum_{i=1}^{12}(\alpha_i - 90)^2$, $\alpha_i$ is the angle of I-Ge-I in the octahedron[3].

| 293 K | | | 373 K | | |
|---|---|---|---|---|---|
| Ge-I bond length (Å) | I-Ge-I bond angle (°) | | Ge-I bond length (Å) | I-Ge-I bond angle (°) | |
| 2.765 | 74.696 | 84.724 | 2.772 | 97.215 | 83.013 |
| 2.792 | 99.189 | 92.789 | 2.772 | 99.463 | 77.76 |
| 2.789 | 82.914 | 94.475 | 3.085 | 94.770 | 95.471 |
| 3.396 | 94.516 | 82.191 | 3.085 | 90.388 | 83.013 |
| 3.485 | 95.104 | 94.416 | 3.523 | 94.770 | 90.388 |
| 3.528 | 82.432 | 100.884 | 3.523 | 77.76 | 97.215 |

**Supplementary Note 2 | More details of ferroelectric characterization.**

Variable-temperature single-crystal X-ray diffraction data were collected on a Rigaku VarimaxTM DW diffractometer with Mo Kα radiation (λ = 0.71073 Å). Data processing with empirical absorption correction was conducted by using the CrysAlisPro 1.171.40.14e (Rigaku OD, 2018). The crystal structures were confirmed by direct methods and refined by full-matrix least-squares methods based on $F^2$ through the OLEX2 and SHELXTL (version 2018) software package. All non-hydrogen atoms were refined anisotropically and all hydrogen atoms were generated geometrically in suitable positions.

The second harmonic generation (SHG) experiments were performed on an Edinburgh FLS 920 by using an unexpanded laser beam with low divergence (pulsed Nd: YAG at a wavelength of 1064 nm, 5 ns pulse duration, 1.6 MW peak power, 10 Hz repetition rate) in the temperature range of 300-380 K. Furthermore, the SHG phase matching tests were performed with a particle size range of 62-375 μm at room temperature (Supplementary Fig. 8).

To derive the non-linear susceptibility, the SHG intensity anisotropy was investigated by using Witec alpha 300 on single-crystal surfaces. SHG measurements were performed in the Imaging Facilities of Molecular Foundry at LBNL. The light coming from a Coherent Chameleon Ultra II Ti: sapphire laser (700 to 1064 nm) with a pulse width of 150 fs and a repetition rate of 50 MHz was applied for the SHG spectroscopic and microscopic measurement. The incident light (frequency ω) was coupled into the microscope using a dichroic mirror (reflective for the light of frequency and transparent for frequency 2ω), then was focused onto the sample with a 100× (0.95 NA) Nikon LU Plan Apo objective and scanned across the sample for measuring SHG in reflection geometry (Supplementary Fig. 9). In the SHG polarimetry measurement, all linear polarization state of input light is available by rotating a half-wave plate (HWP) in the laser beam path. Before the light passes through HWP, there is also a polarizer to enhance the degree of linear polarization. The HWP was mounted on a motorized rotation stage and controlled by the software, allowing for a continuous and precise rotation of incident polarization $E$ (φ) in the $x$-$y$ plane of the laboratory coordinate system ($x$, $y$, $z$), where φ is the azimuthal angle of the fundamental light polarization.

For the dielectric measurements, the sample of $DMAGeI_3$ was made with single-crystals cut perpendicular to the *a*-, *b*-, and *c*-axis respectively. Silver conduction paste deposited on the plate surfaces was used as the electrodes. The temperature-dependent dielectric constants were performed on the Tonghui TH2828A instrument under the frequency range from 1 kHz to 1 MHz with an applied electric field of 1 V (Supplementary Fig. 11).

Single crystal electrodes of $DMAGeI_3$ were prepared for the *P*-*E* hysteresis loop measurements by Sawyer-Tower method. To reduce the coercive force field, single crystals obtained were processed to make crystals along the polar axis (*c*-axis). Then a pair of electrodes were formed orthogonally with each other on a single crystal of 1 with silver conductive adhesive, and copper wires were used to connect with IC sockets (Supplementary Fig. 12). Ferroelectric polarization imaging and local switching studies were carried out using a resonant-enhanced PFM (MFP-3D, Asylum Research). To visualize the domain switching process, a local polarization manipulation study on the bulk crystal surface was carried out by applying a DC tip bias of ±50 V for 0.5 s. The polarization in some domains can be switched forth and back in a reproducible manner. The macroscopic piezoelectric coefficient ($d_{33}$) was measured by a commercial piezometer (Piezotest, model: PM200) using the quasi-static method (Supplementary Fig. 13).

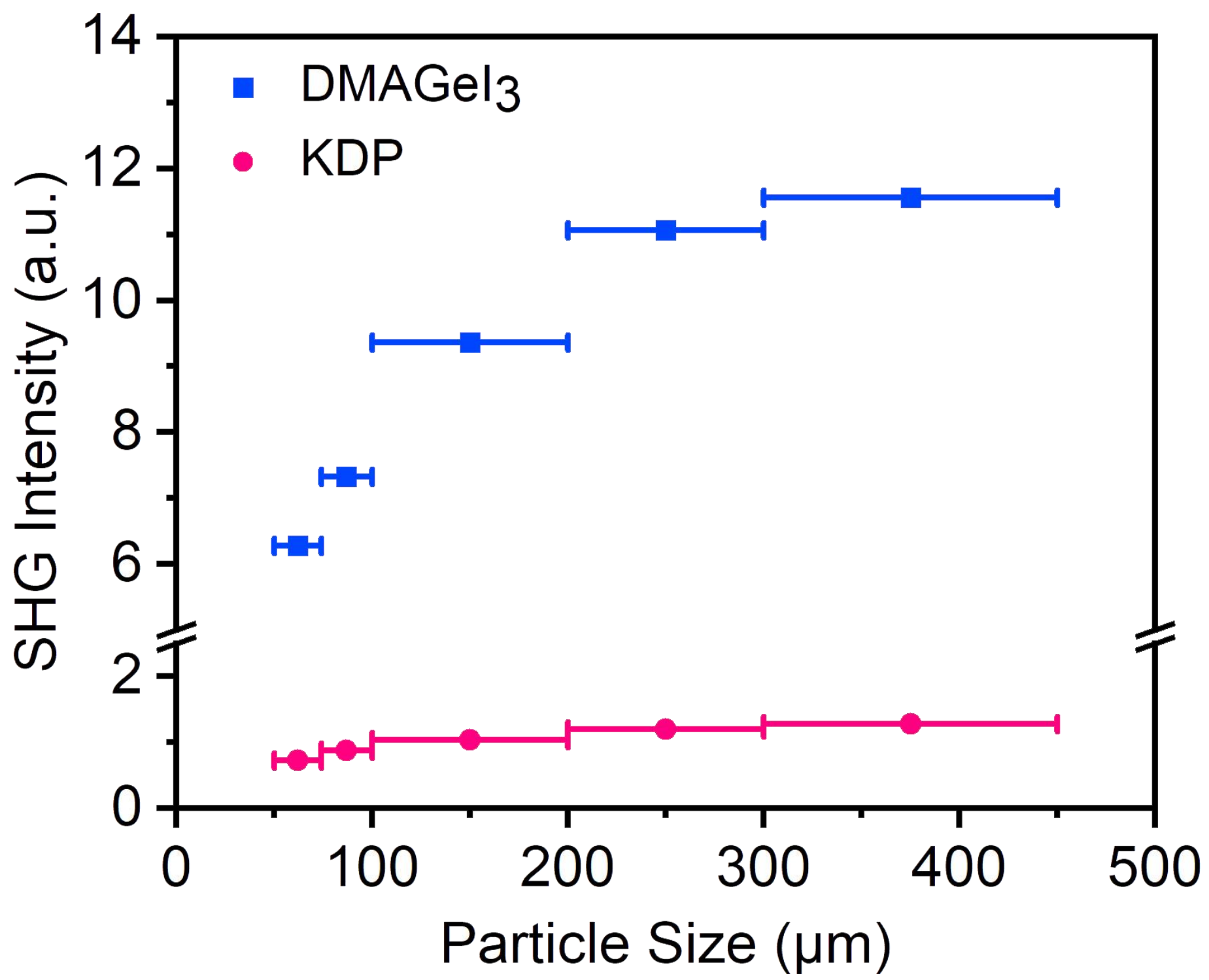

**Supplementary Fig. 8 | The particle size (diameter) dependence of SHG intensity for both DMAGeI₃ and KDP under 1064 nm laser radiation.** The SHG intensities increase with increasing particle sizes until it attains the constant, indicating the phase matchable nature of the DMAGeI₃.

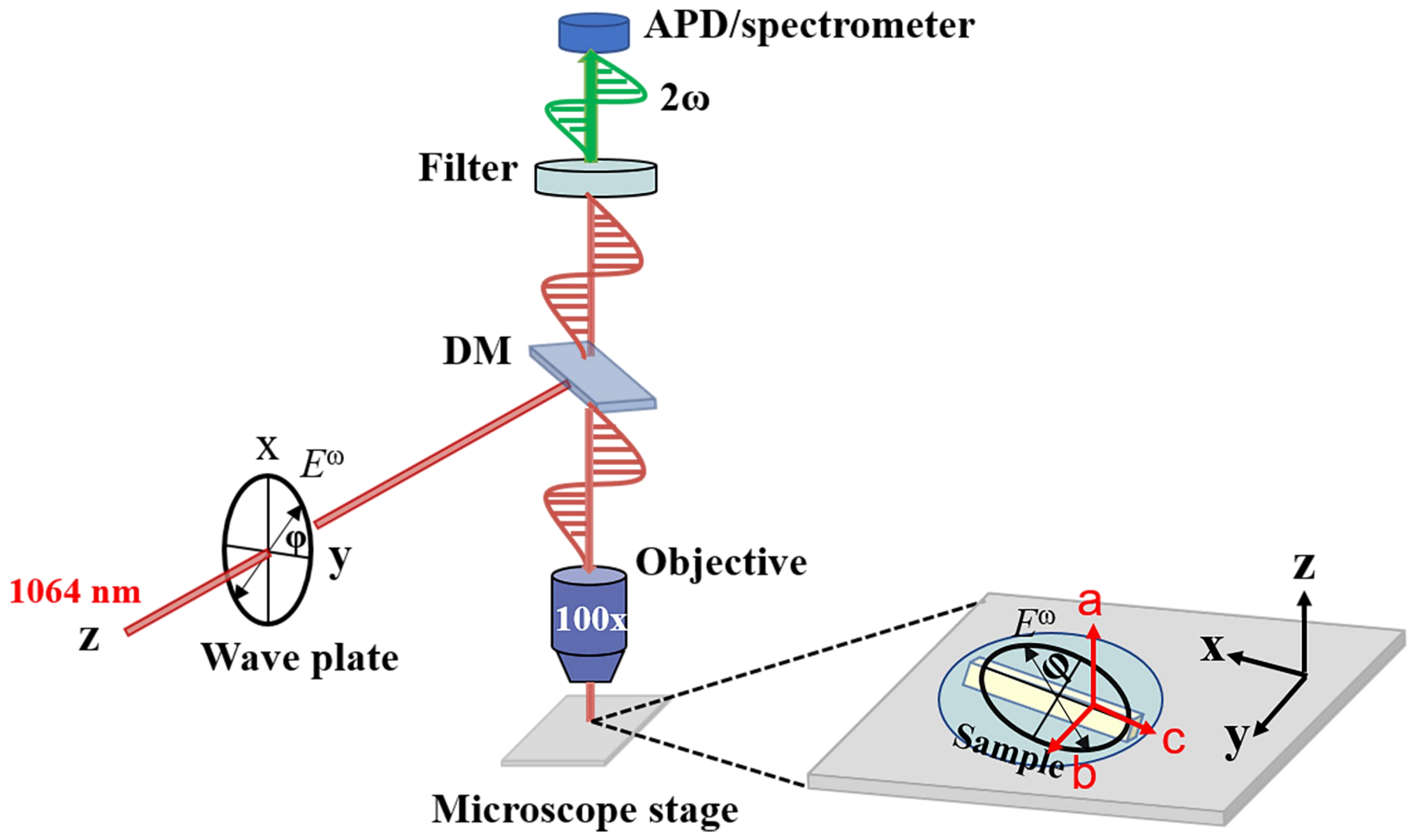

**Supplementary Fig. 9 | Schematic of the experimental setup to illustrate the sample orientation with respect to the stage geometry and light propagation direction.**

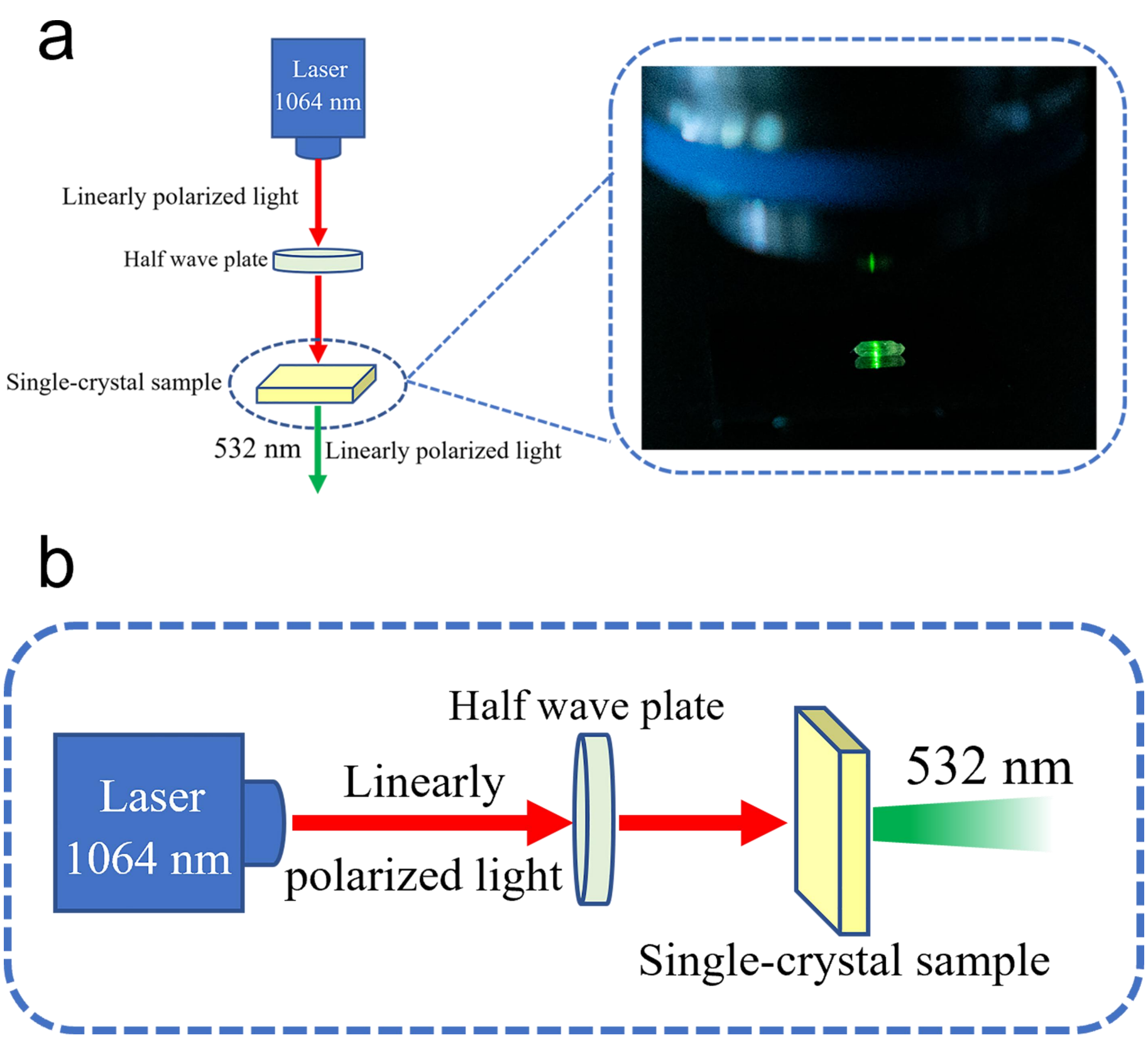

**Supplementary Fig. 10 | a, Experimental diagram and photograph of crystal. b, Schematic of the non-linear optical converter.** The bright green light can be clearly observed on the crystal of $DMAGeI_3$ under 1064 nm linearly polarized light, which indicates that the sample generates strong second-harmonic light with wavelength of 532 nm.

**Supplementary Table 4 | Comparison of SHG signal strength between reported well known inorganic nonlinear optical material and organic-inorganic hybrid perovskite and $DMAGeI_3$ in this work.** BBO: $\beta$-$BaB_2O$; LBO: $LiB_3O_5$; KABO: $K_2Al_2B_2O$; LN: $LiNbO_3$; DM: dimethylammonium; PM: phenylmethanaminium; 3-dcbm: 3,3-difluorocyclobutylammonium; 2-fbm: 2-fluorobenzylammonium; R-EQ: (R)-N-ethyl-3-quinuclidinol; DFCBA: 3,3-Difluorocyclobutylammonium; 3-HQCM: R-N-methyl-3-hydroxylquinuclidinium; *S*-1-CEM : *S*-1-(4-chlorophenyl)ethylammonium; NPD: *N*-methylpyrrolidinium; R3HQ: *R*-3-hydroxylquinuclidinium; EATMP: (2-aminoethyl)trimethylphosphanium]; TMIM: (CH3)3NCH2I; SPAD: 1-((2-hydroxybenzylidene)amino)pyridin-1-ium); MHy: $CH_3NH_2NH_2$; MA: CH3NH3; FA: CH(NH2)2; TMAEA: 2-trimethylammonioethylammonium

| | Compound | SHG intensity times vs $KH_2PO_4$ (KDP) | Reference |
|---|---|---|---|
| well-known inorganic nonlinear optical material | BBO | ~4.4* | 4-7 |
| | LBO | ~2.4* | 5,7-8 |
| | KABO | ~1.34* | 5 |
| | LN | ~12* | 5 |
| Organic-inorganic | $[3\text{-dcbm}]_2CuCl_4$ | ~0.25 | 9 |
| hybrid perovskite | $[2\text{-fbm}]_2PbCl_4$ | ~0.9 | 10 |
| | $[R\text{-EQ}]PbI_3$ | 0.21 | 11 |
| | $[DFCBA]_2CrCl_4$ | 0.25 | 12 |
| | $(3\text{-HQCM})_2RbCe(NO_3)_6$ | ~0.25 | 13 |
| | $(S\text{-1-CEM})_2PbI_4$ | ~0.5 | 14 |
| | $(NPD)_3Sb_2Br_9$ | ~0.6 | 15 |
| | $(R_3HQ)_4KCe(NO_3)_8$ | ~0.55 | 16 |
| | $(C_4H_9NH_3)_2(NH_3CH_3)_2Sn_3Br_{10}$ | ~1 | 17 |
| | $[(CH_3)_2CHCH_2NH_3]_2PbCl_4$ | ~1 | 18 |
| | $(EATMP)Pb_2Br_6$ | ~0.17 | 19 |
| | $(TMIM)PbI_3$ | ~0.65 | 20 |
| | $[CH_3(CH_2)_3NH_3]_2(CH_3NH_3)Pb_2Br_7$ | ~0.4 | 21 |
| | $(C_6H_5CH_2NH_3)_2CsAgBiBr_7$ | ~1 | 22 |
| | $[SAPD]PbI_3$ | ~2 | 23 |
| | $MHyPbCl_3$ | ~0.03 | 24 |
| | $MHyPbBr_3$ | ~0.18 | 25 |
| | $NH(CH_3)_3SnCl_3$ | ~1 | 26 |
| | $NH(CH_3)_3SnBr_3$ | ~2.5 | 26 |
| | $MAGeBr_3$ | ~5.3 | 28 |
| | $FAGeBr_3$ | ~0.9 | 28 |
| | $FA_{0.5}MA_{0.5}GeBr_3$ | ~1.95 | 28 |
| | $[TMAEA]Pb_2Cl_6$ | ~0.2 | 29 |
| | $(4,4\text{-DFM})_4AgBiI_8$ | ~1 | 30 |
| | $DMAGeI_3$ | ~12 | this work |

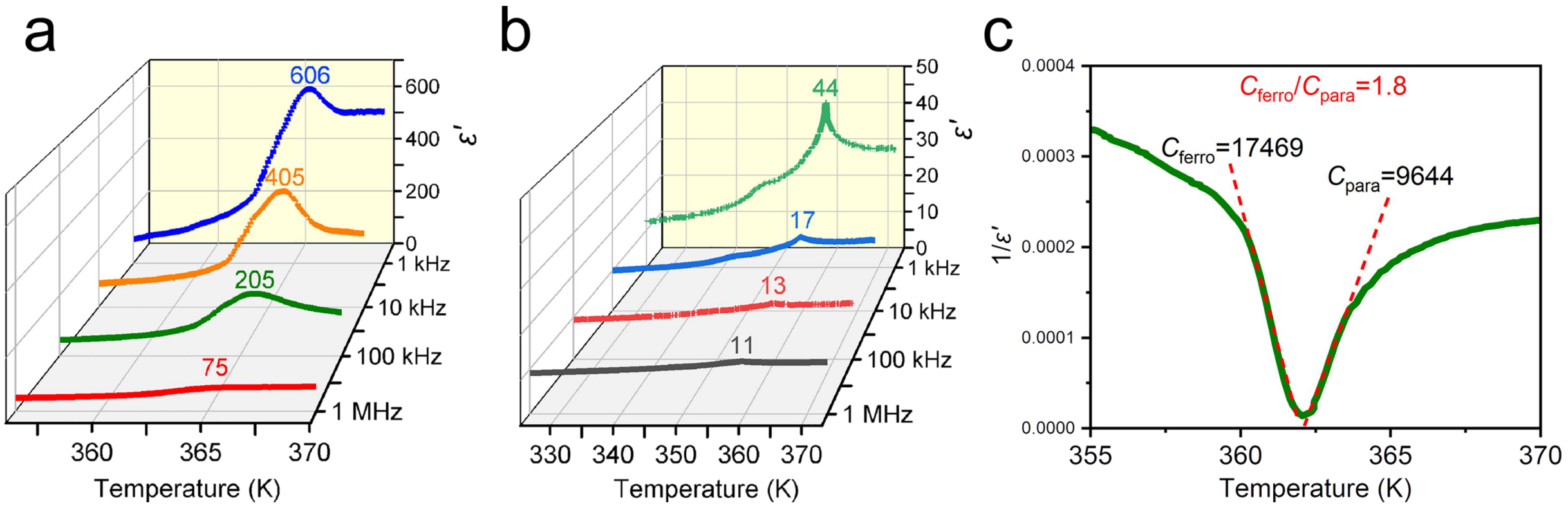

**Supplementary Fig. 11 | Temperature-dependent dielectric real part ($\varepsilon'$) of $DMAGeI_3$ measured at various frequencies.** **a**, along the *a*-axis. **b**, along the *b*-axis. **c**, The fitting to Curie-Weiss law of dielectric anomalies in the vicinity of $T_c$ at 1 kHz.

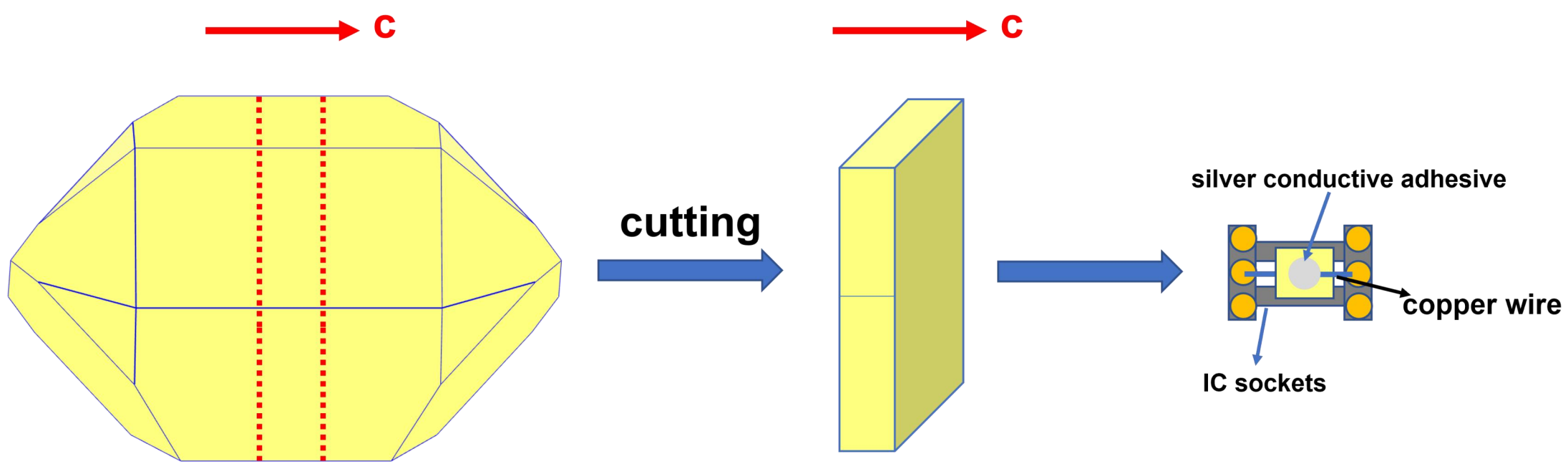

**Supplementary Fig. 12 | Crystal process of $DMAGeI_3$, cut perpendicular to the *c*-axis.**

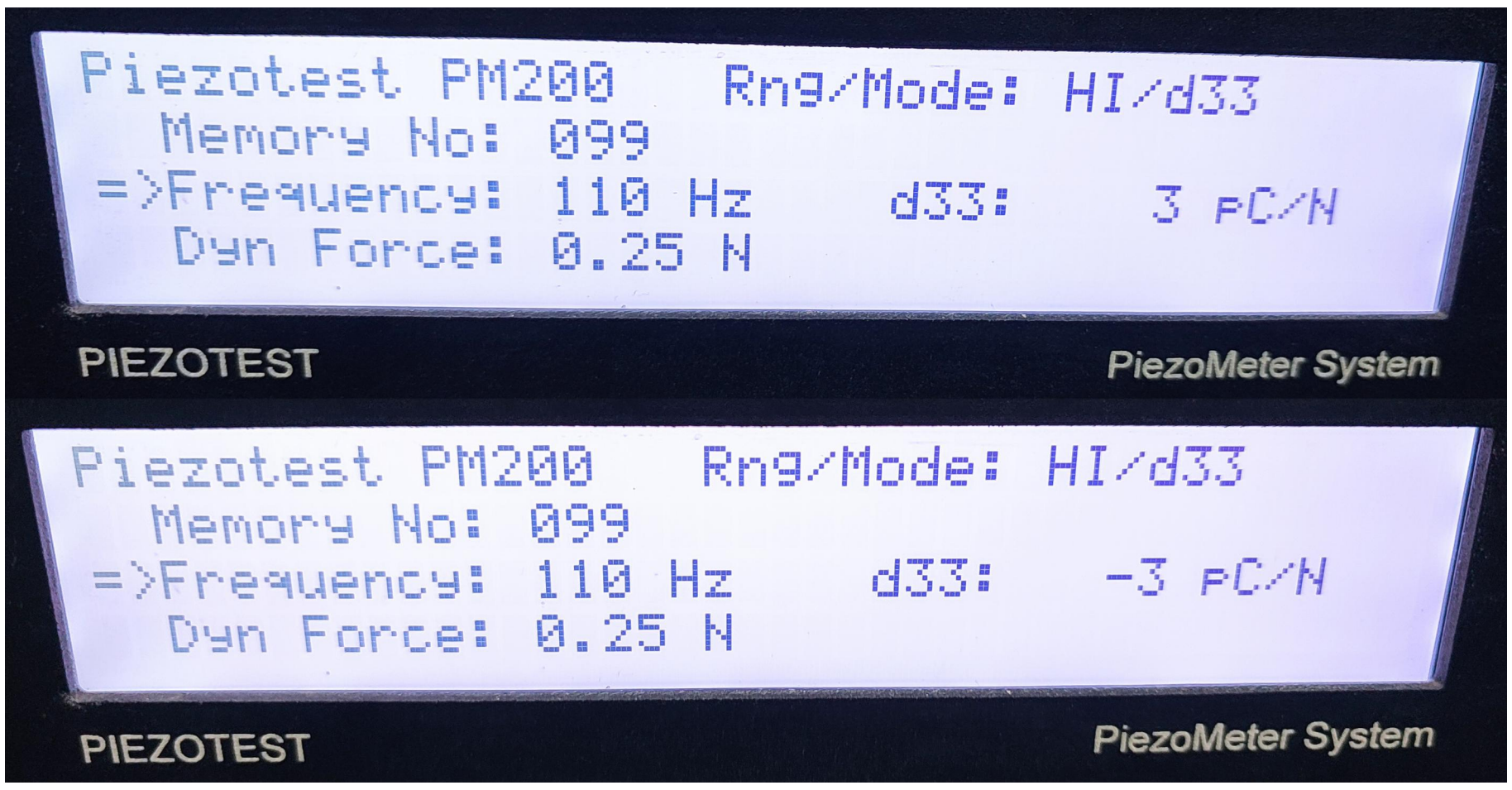

**Supplementary Fig. 13 | Photos of $d_{33}$ meter.** The maximum $d_{33}$ of $DMAGeI_3$ is along the vicinity of the [001] direction of the crystal.

**Supplementary Table 5 | Comparison of the maximum dielectric real part ($\varepsilon'$) at $T_c$ of organic-inorganic hybrid perovskite ferroelectrics and $DMAGeI_3$ in this work.** BM: benzylammonium; R3HQ: *R*-3-hydroxylquinuclidinium; NPD: *N*-methylpyrrolidinium; 4,4-DCM: 4,4-difluorocyclohexylammonium; RM3HQ: *R*-*N*-methyl-3-hydroxylquinuclidinium; MDABCO: *N*-methyl-*N'*-diazabicyclo[2.2.2]octonium; 4,4-DFM: 4,4-difluoropiperidinium; TMCM: trimethylchloromethyl ammonium.

| Compound | Dielectric | Reference |
|---|---|---|
| $(BM)_2PbCl_4$ | ~900 (0.5 kHz) | 11 |
| $(R3HQ)_4KCe(NO_3)_8$ | ~135(1 kHz) | 16 |
| $(NPD)_3Sb_2Br_9$ | ~460 (1 kHz) | 15 |
| $(4,4\text{-}DCM)_2PbI_4$ | ~90 (0.5 kHz) | 32 |
| $(RM3HQ)_2RbLa(NO_3)_6$ | ~23 (2 kHz) | 33 |
| $(C_4H_9NH_3)_2PbCl_4$ | ~65 (1 kHz) | 34 |
| MDABCO–$NH_4I_3$ | ~15000 (20 Hz) | 35 |
| (3-Pyrrolinium)($CdCl_3$) | ~4600 (0.5 kHz) | 36 |
| $(4,4\text{-DFPD})_2PbI_4$ | ~2200 (0.5 kHz) | 37 |
| $[(CH_3)_2(F\text{-}CH_2CH_2)NH]_3(CdCl_3)(CdCl_4)$ | ~176 (1 kHz) | 38 |
| (S)-3-F-(pyrrolidinium)$CdCl_3$ | ~1500 (0.5 kHz) | 39 |
| (4,4-difluoropiperidinium)$_4$$AgBiI_8$ | ~1700 (0.5 kHz) | 30 |
| TMCM-$MnCl_3$ | ~1200 (1 kHz) | 40 |
| $BaTiO_3$ | 10000 | 41 |
| KDP | 20000 | 41 |
| TGS | 2000 | 42 |
| Rochelle salt | 4000 | 41 |
| PVDF | 50 | 41 |
| Croconic acid | 850 (30 kHz) | 41 |
| DMAGe$I_3$ | 102988(1 kHz) | This work |

**Supplementary Table 6 | Comparison of spontaneous polarization $P_s$ in a variety of materials and $DMAGeI_3$ in this work.** TGS: triglycine sulphate; PVDF: Polyvinylidene Fluoride; MDABCO: *N*-methyl-*N′*-diazabicyclo[2.2.2]octonium; QP: Quinuclidinium perrhenate TMCM: trimethylchloromethyl ammonium; 4,4-DFM: 4,4-difluoropiperidinium; EATMP: (2-aminoethyl)trimethylphosphanium; dabco: diazabicyclo[2.2.2]octane.

| Compound | $P_s$ (μC/cm$^2$) | Reference |
|---|---|---|
| TGS | 3.8 | 41 |
| $BaTiO_3$ | 27 | 42 |
| PVDF | 8 | 42 |
| $CsGeI_3$ | 20 | 43 |
| Croconic acid | 20 | 42 |
| MDABCO–$NH_4I_3$ | 22 | 35 |
| TMCM-$MnCl_3$ | 4 | 40 |
| TMCM-$CdCl_3$ | 6 | 40 |
| QP | 4.5 | 44 |
| Rochelle salt | 0.25 | 42 |
| $[NH_4][Zn(HCOO)_3]$ | 0.68 | 45 |
| (3-bromopropylammonium)$_2PbBr_4$ | 4.8 | 46 |
| (4,4-DFPD)$_2PbI_4$ | 10 | 37 |
| $[CH_3(CH_2)_3NH_3]_2(CH_3NH_3)Pb_2Br_7$ | 3.6 | 21 |
| $(C_4H_9NH_3)_2(C_2H_5NH_3)_2Pb_3Br_{10}$ | 5 | 47 |
| $(C_4H_9NH_3)_2(CH_3NH_3)_2Pb_3Br_{10}$ | 2.9 | 48 |
| (EATMP)$PbBr_4$ | 0.95 | 49 |
| [Hdabco]$BF_4$ | 4.9 | 50 |
| [Hdabco]$ClO_4$ | 4 | 51 |
| diisopropylammonium bromide | 23 | 52 |
| $[Me_3NCH_2CH_2OH]CdCl_3$ | 17.1 | 53 |
| $DMAGeI_3$ | 24.14 | This work |

**Supplementary Table 7 | Comparison of coercive field $E_c$ in a variety of materials and $DMAGeI_3$ in this work**. TGS: triglycine sulphate; PVDF: Polyvinylidene Fluoride; MDABCO: *N*-methyl-*N′*-diazabicyclo[2.2.2]octonium; TMCM: trimethylchloromethyl ammonium; 4,4-DFM: 4,4-difluoropiperidinium; QP: Quinuclidinium perrhenate; EATMP: (2-aminoethyl)trimethylphosphanium; dabco: diazabicyclo[2.2.2]octane

| Compound | $E_c$ (kV/cm) | Reference |
|---|---|---|
| TGS | 0.9 | 41 |
| $BaTiO_3$ | 10 | 42 |
| PVDF | 500 | 42 |
| $CsGeI_3$ | ~40 | 43 |
| Croconic acid | 14 | 42 |
| MDABCO–$NH_4I_3$ | 12 | 35 |
| TMCM-$MnCl_3$ | ~27.5 | 40 |
| TMCM-$CdCl_3$ | ~9 | 40 |
| QP | ~4 | 44 |
| Rochelle salt | 0.2 | 42 |
| $[NH_4][Zn(HCOO)_3]$ | 2.8 | 45 |
| (3-bromopropylammonium)$_2PbBr_4$ | 18.2 | 46 |
| (4,4-DFPD)$_2PbI_4$ | 7.1 | 37 |
| $[CH_3(CH_2)_3NH_3]_2(CH_3NH_3)Pb_2Br_7$ | 26 | 21 |
| $(C_4H_9NH_3)_2(C_2H_5NH_3)_2Pb_3Br_{10}$ | 8 | 47 |
| $(C_4H_9NH_3)_2(CH_3NH_3)_2Pb_3Br_{10}$ | 16 | 48 |
| (EATMP)$PbBr_4$ | 7.3 | 49 |
| [Hdabco]$BF_4$ | 230 | 50 |
| [Hdabco]$ClO_4$ | 83 | 51 |
| $DMAGeI_3$ | 0.83-2.2 | This work |

**Appendix S4.** Acronyms arranged in top-down order in **Supplementary Table 7**.
TGS: triglycine sulphate; PVDF: Polyvinylidene Fluoride;
MDABCO: *N*-methyl-*N′*-diazabicyclo[2.2.2]octonium;
TMCM: trimethylchloromethyl ammonium; 4,4-DFM: 4,4-difluoropiperidinium;
QP: Quinuclidinium perrhenate;
EATMP: (2-aminoethyl)trimethylphosphanium; dabco: diazabicyclo[2.2.2]octane

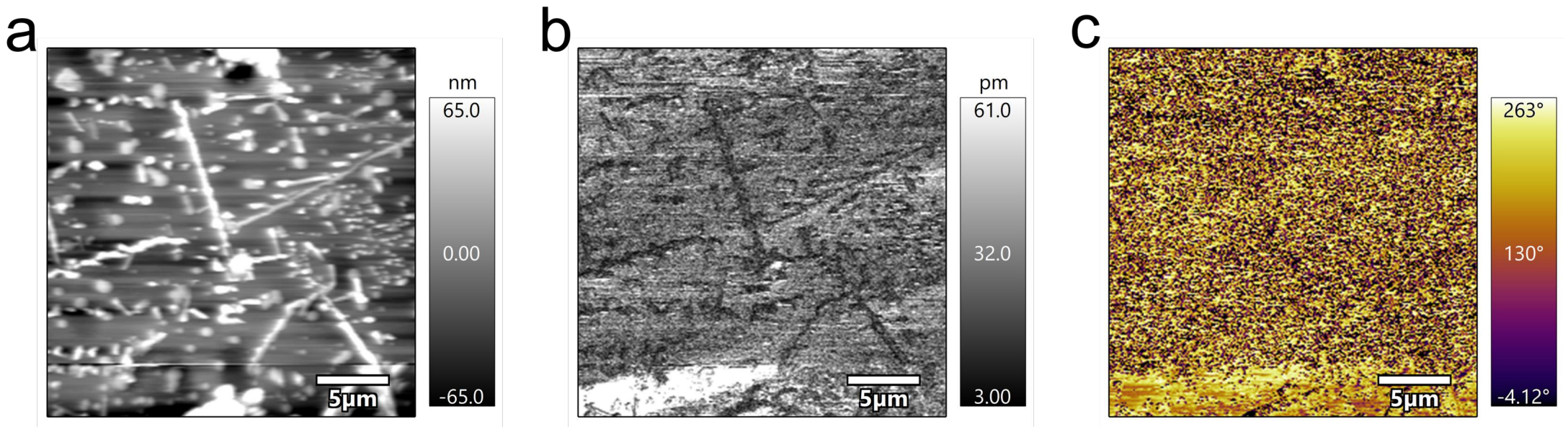


**Supplementary Fig. 14 | Domain structures in the (010) plane of $DMAGeI_3$. a**, Topography. **b**, Vertical amplitude images. **c**, The corresponding vertical phase images.

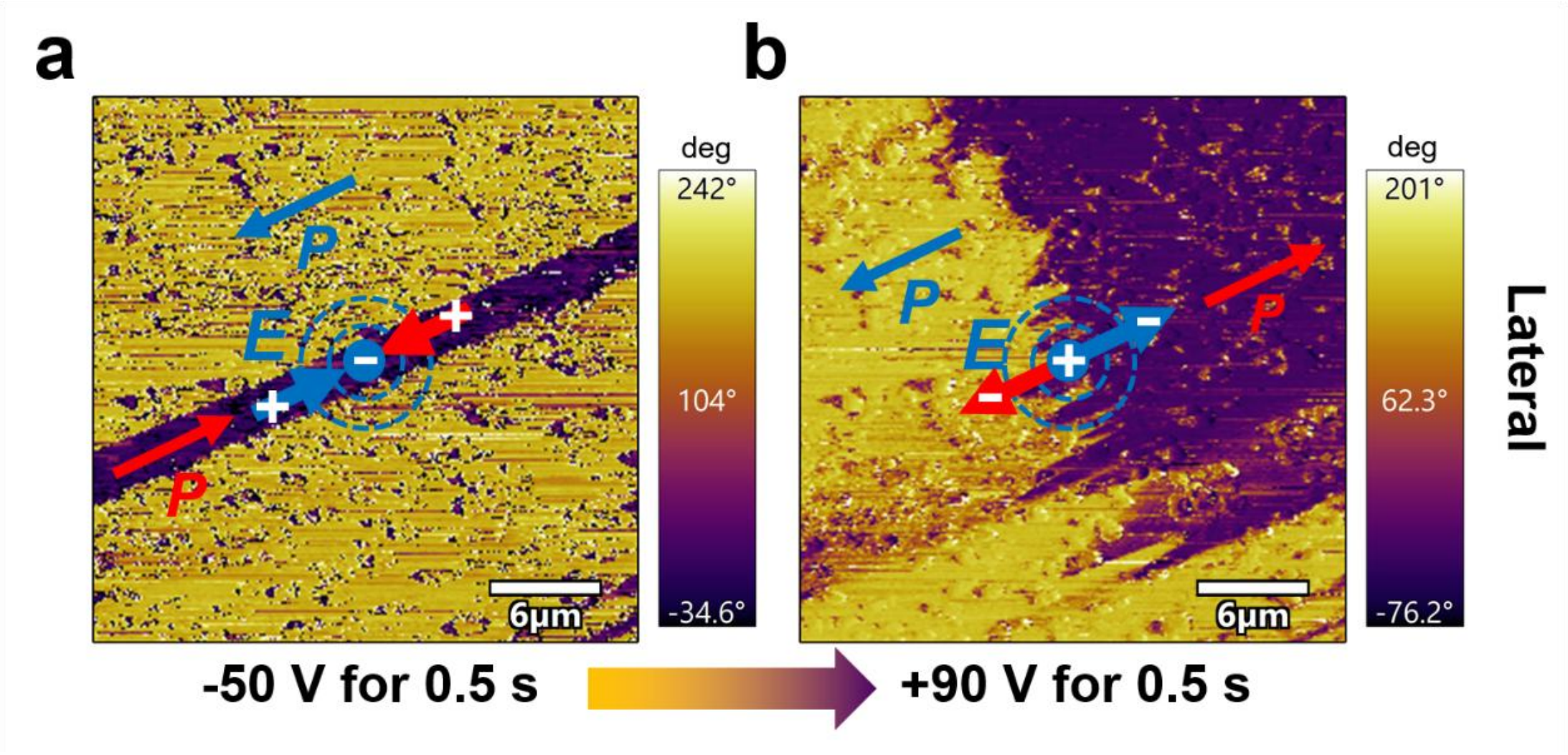


**Supplementary Fig. 15 | The measurements of polarization reversal for DMAGeI$_3$ in lateral PFM. a**, Phase image the same as Fig. 3i. **b**, Succeeding phase image after applying positive bias of +90 V for 0.5 s.

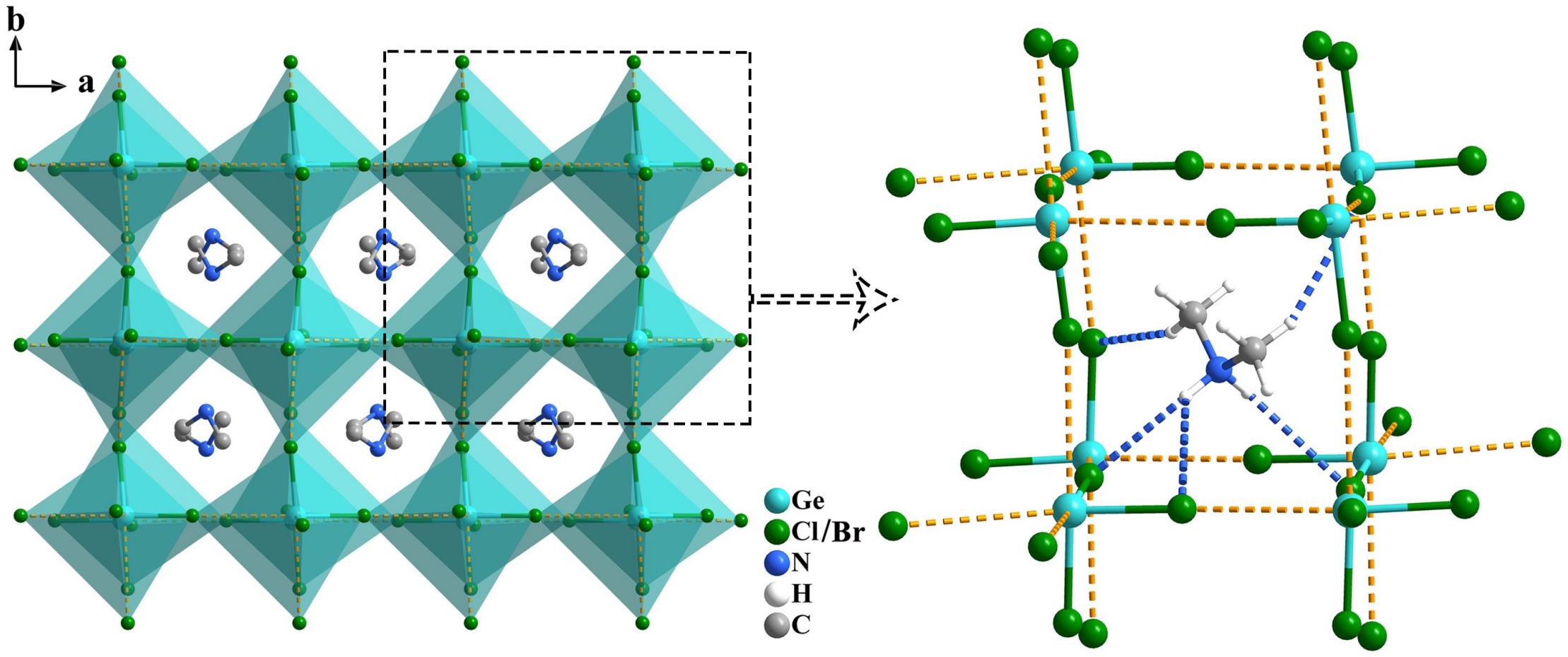


**Supplementary Fig. 16 | Structural illustration of the 3D perovskite $DMAGeCl_3$ and $DMAGeBr_3$ viewed along the *b*-axis.** The isostructural compounds $DMAGeCl_3$ and $DMAGeBr_3$ crystallize in the orthorhombic system of space group *Pbca* at room temperature. The $DMA^+$ are antiparallelly aligned along with the *b*-axis through the hydrogen bonds interaction with the perovskite cage, which causes the disappearance of the molecular polarization. More detailed structural information is summarized in Supplementary Tables 8-9.

**Supplementary Table 8 | Crystal data and structure refinement details compound $DMAGeCl_3$ and $DMAGeBr_3$.**

| Compound | $DMAGeCl_3$ | $DMAGeBr_3$ |
|---|---|---|
| Temperature/K | 293 | 293 K |
| Empirical formula | $C_2H_8NGeCl_3$ | $C_2H_8NGeBr_3$ |
| Formula weight | 225.05 | 358.40 |
| Crystal system | orthorhombic | orthorhombic |
| Space group | *Pbca* | *Pbca* |
| a/Å | 11.5752(8) | 11.9011(8) |
| b/Å | 11.8408(8) | 12.2582(7) |
| c/Å | 12.0003(10) | 12.2854(8) |
| $\alpha$/° | 90 | 90 |
| $\beta$/° | 90 | 90 |
| $\gamma$/° | 90 | 90 |
| Volume/Å$^3$ | 1644.8(2) | 1792.3(2) |
| Z | 8 | 8 |
| *F* (000) | 880.0 | 1312.0 |
| Radiation | Mo Kα | Mo Kα |
| absorption correction | emiempirical | emiempirical |
| GOF | 1.039 | 1.007 |
| $R_1$ | 0.0441 | 0.0651 |
| $wR_2$ | 0.1169 | 0.1713 |

**Supplementary Table 9 | Selected bond lengths [Å] and angles [°] for compounds**

**$DMAGeCl_3$ and $DMAGeBr_3$ at 293 K, respectively.**

| $DMAGeCl_3$ | | $DMAGeBr_3$ | |
|---|---|---|---|
| Ge01—Cl02 | 2.3249 (11) | Ge01—Br02 | 2.4876 (16) |
| Ge01—Cl03 | 2.3150 (12) | Ge01—Br03 | 2.4676 (18) |
| Ge01—Cl04 | 2.3454 (12) | Ge01—Br04 | 2.5001 (17) |
| N005—C006 | 1.456 (4) | N005—C006 | 1.434 (14) |
| N005—C007 | 1.461 (5) | N005—C1 | 1.456 (16) |
| | | | |
| Cl02—Ge01—Cl04 | 93.76 (4) | Br02—Ge01—Br04 | 94.64 (6) |
| Cl03—Ge01—Cl02 | 94.51 (3) | Br03—Ge01—Br02 | 95.47 (6) |
| Cl03—Ge01—Cl04 | 92.16 (5) | Br03—Ge01—Br04 | 93.65 (6) |
| C006—N005—C007 | 115.7 (3) | C006—N005—C1 | 116.1 (11) |

**Supplementary Note 3 | More details of DFT calculations.**

For comparison, the PBE functional modified for solids (PBEsol) and the D2 Grimme correction for van der Waals interactions have also been tested [54, 55]. The test results of optimized lattice constants indicate that PBE+D3 leads to the best agreement with the experimental data, as compared in Supplementary Fig. 17. Thus, the PBE+D3 is adopted as the default choice in our DFT calculations.

To partition the individual ferroelectric contribution from the $GeI_3$ framework and DMA cation, two hypothetic structures are considered. Based on the optimized $DMAGeI_3$, the pure $GeI_3$ framework is obtained by removing all DMA cations. To keep the charge neutrality, one electron/f.u. is added manually to the $GeI_3$ framework. No further structural optimization is performed. Similar process can be done to obtain pure DMA group by removing the $GeI_3$ framework and deducting electrons correspondingly. Then their corresponding polarization can be calculated. The direct adding of these two individual contributions is 29.80 μC/cm$^2$, very close to the global polarization 29.09 μC/cm$^2$. The tiny bias may originate from the interaction between DMA and $GeI_3$ parts, which is negligible here.

Furthermore, as presented in Supplementary Fig. 19, the electron density difference, defined as $\triangle\rho=\rho(DMAGeI_3)-\rho(DMA)$ where $\rho(DMAGeI_3)$ and $\rho(DMA)$ are the electron density of $DMAGeI_3$ and DMA with adding hole/f.u., is calculated to support the rationality. It is obvious that the residual electrons only locate at the $GeI_3$ framework. Similar situation occurs for $\triangle\rho=\rho(DMAGeI_3)-\rho(GeI_3)$ where $\rho(GeI_3)$ are the electron density of $GeI_3$ framework with adding electron/f.u.: the residual electrons only locate at the DMA group. Therefore, our treatment of

individual parts can restore the charge distribution in original $DMAGeI_3$.

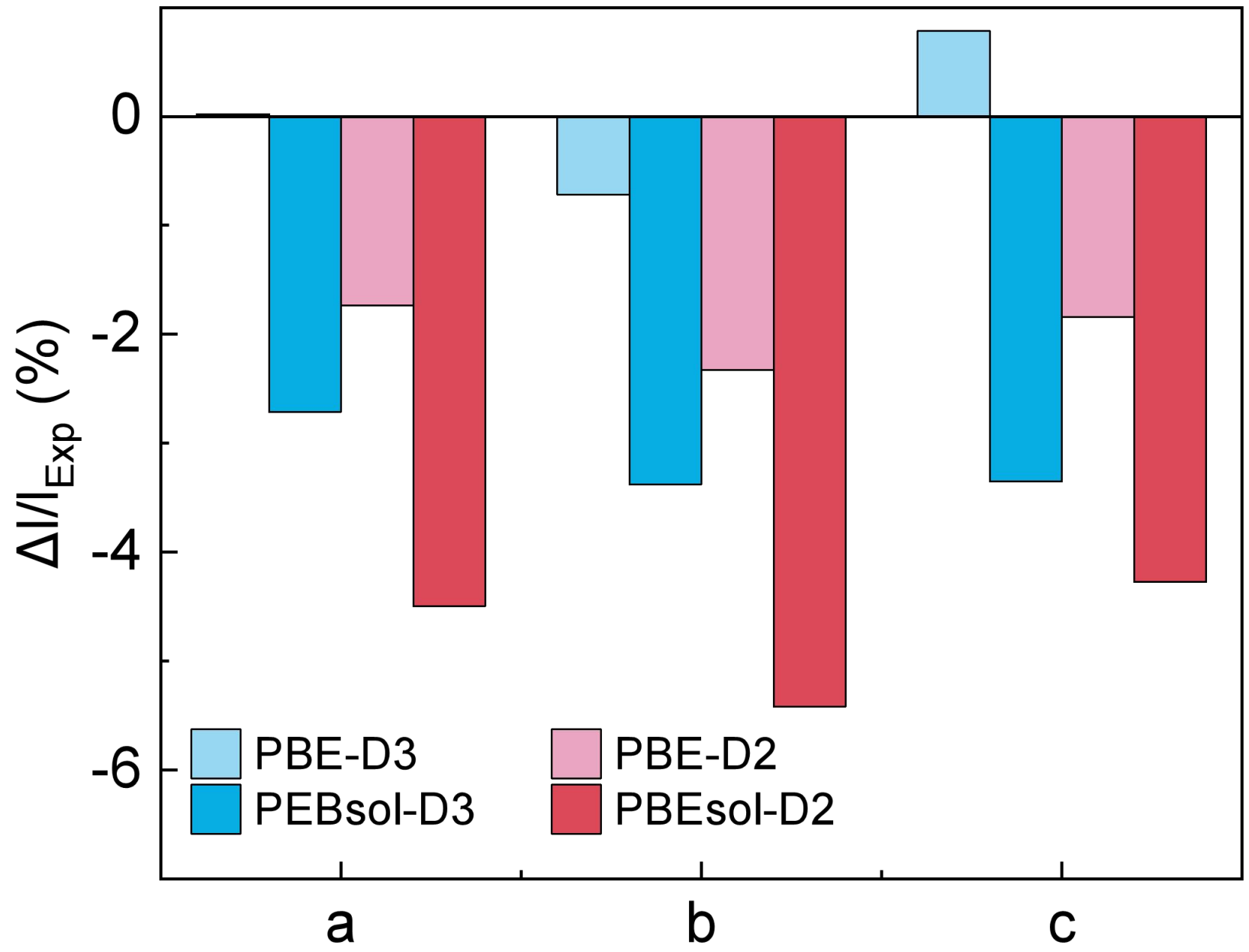


**Supplementary Fig. 17 | The test of functionals and van der Waals corrections.** $\triangle l = l - l_{exp}$, where $l$ and $l_{Exp}$ are the lattice constants obtained from DFT calculation and experimental data.

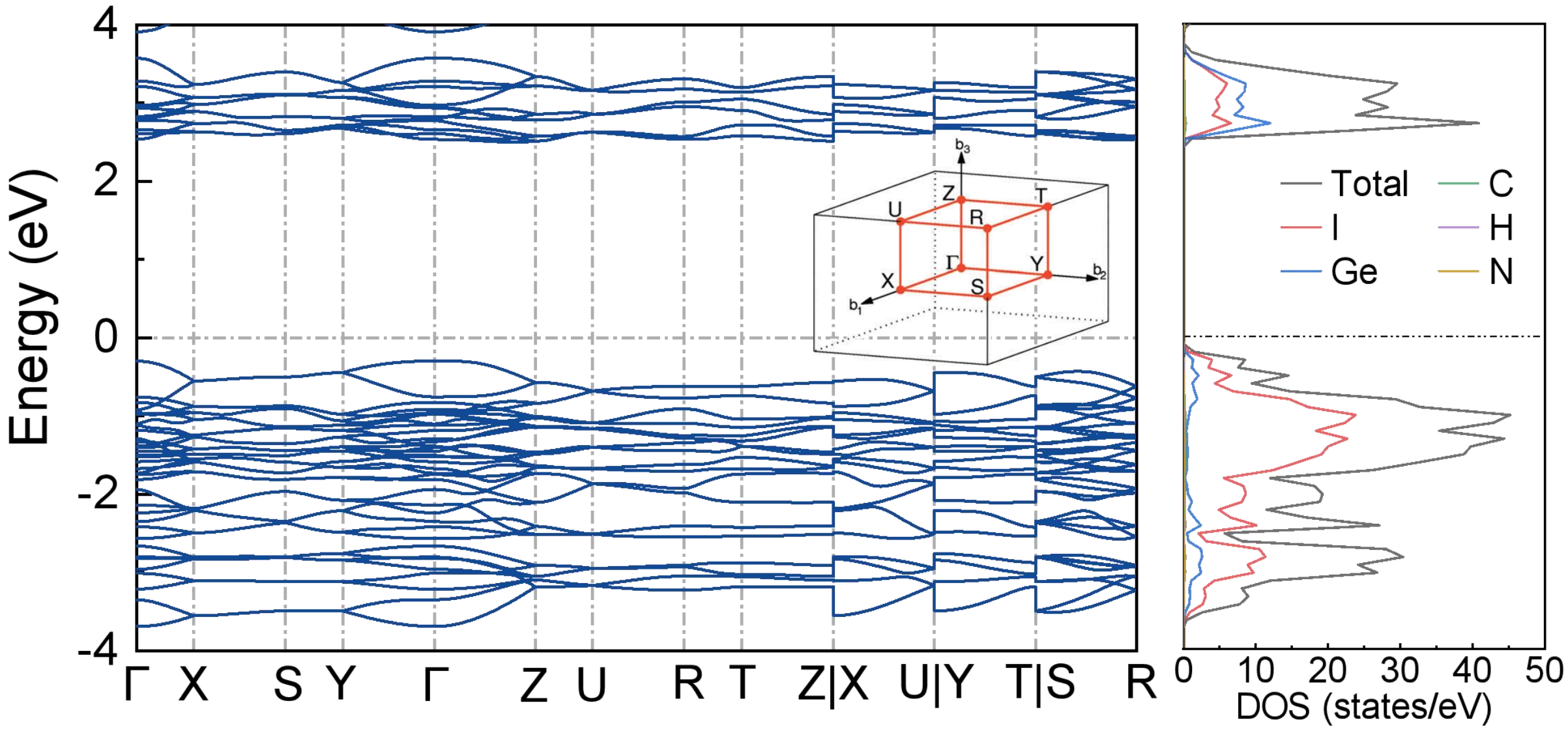


**Supplementary Fig. 18 | DFT electronics properties of $DMAGeI_3$.** Left: band structure. Inset: the schematic of Brillouin zone and $k$-path for the $DMAGeI_3$. The direct band gap at Γ point is 2.79 eV. Right: atom-projected DOS. The bands near Fermi level are mainly contributed by the $GeI_3$ framework.

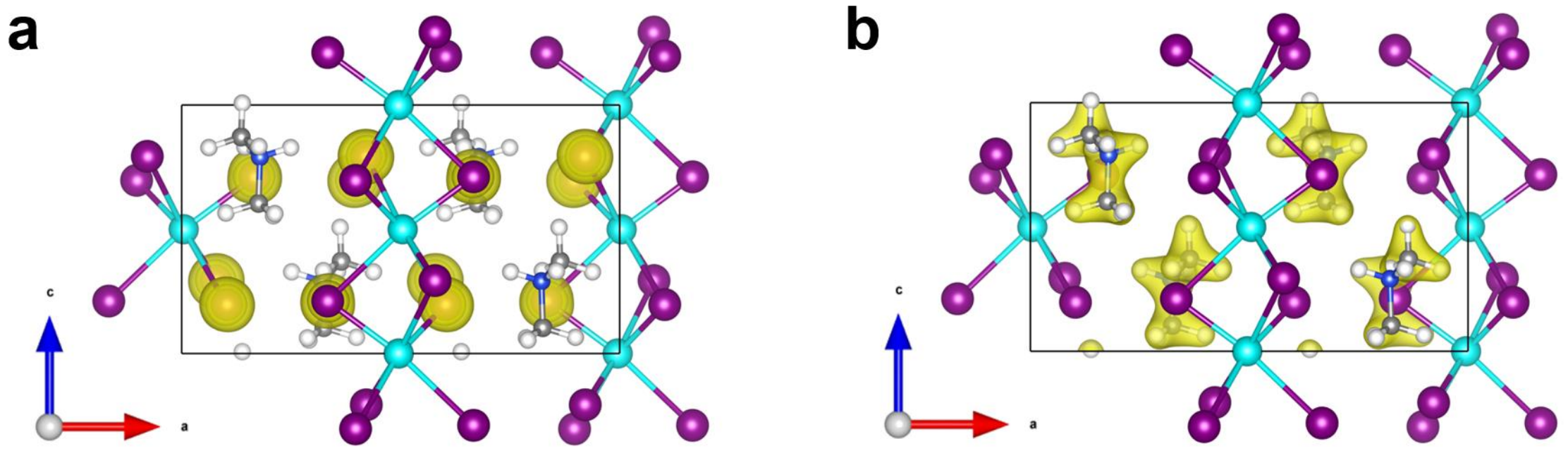


**Supplementary Fig. 19 | Electron density difference between $DMAGeI_3$ and extracted parts with holes/electrons added. a,** Between the $DMAGeI_3$ and the pure DMA group. **b,** Between the $DMAGeI_3$ and the $GeI_3$ framework. The isosurfaces are set to 0.1 $e$/Å$^3$.

**Supplementary Note 4 | More details of SHG calculations.**

The SHG intensity can be estimated via $I \propto (P^{2\omega})^2 = P_a^2 + P_b^2 + P_c^2$, where $P^2$ is the second harmonic polarization generated by the electric field of incident light with angular frequency. $a/b/c$ are perpendicular crystalline axes here. For the LTP $DMAGeI_3$ with $Pna2_1$ space group, $P^{2\omega}$ can be given by [56]:

$$\begin{pmatrix} P_a \\ P_b \\ P_c \end{pmatrix} = 2\times\begin{pmatrix} 0 & 0 & 0 & 0 & d_{15} & 0 \\ 0 & 0 & 0 & d_{24} & 0 & 0 \\ d_{31} & d_{32} & d_{33} & 0 & 0 & 0 \end{pmatrix}\begin{pmatrix} E_a^2 \\ E_b^2 \\ E_c^2 \\ 2E_bE_c \\ 2E_aE_c \\ 2E_aE_b \end{pmatrix} = 2\times\begin{pmatrix} 2d_{15}E_aE_c \\ 2d_{24}E_bE_c \\ d_{31}E_a{}^2 + d_{32}E_b{}^2 + d_{33}E_c{}^2 \end{pmatrix}, \quad (1)$$

where $d_{ij}$'s are the elements of second-harmonic nonlinear optical susceptibility tensor, and $E_i$ is component of electric field of incident light.

As presented in Supplementary Fig. 20a, we consider the $bc$ plane which is the most possible one in our experimental setup. Then the propagation direction of incident light is parallel to the $a$ axis. The electric field of incident light can be written as $E$=($E_a$, $E_b$, $E_c$)=$E$(0, cos$\varphi$, sin$\varphi$), where $j$ is the angle between the $b$-axis and electric field. Then, the SHG intensity of $DMAGeI_3$ is:

$$I_{DMAGeI_3} \propto (P^{2\omega})^2 = P_a^2 + P_b^2 + P_c^2 = (2d_{24}\sin 2\varphi)^2 + (2d_{32}\cos^2\varphi + 2d_{33}\sin^2\varphi)^2. \quad (2)$$

For KDP with $I\bar{4}2d$ space group, $P^{2\omega}$ can be given by [56]:

$$\begin{pmatrix} P_a \\ P_b \\ P_c \end{pmatrix} = 2\times\begin{pmatrix} 0 & 0 & 0 & d_{14} & 0 & 0 \\ 0 & 0 & 0 & 0 & d_{14} & 0 \\ 0 & 0 & 0 & 0 & 0 & d_{36} \end{pmatrix}\begin{pmatrix} E_a^2 \\ E_b^2 \\ E_c^2 \\ 2E_bE_c \\ 2E_aE_c \\ 2E_aE_b \end{pmatrix} = 2\times\begin{pmatrix} 2d_{14}E_bE_c \\ 2d_{14}E_aE_c \\ 2d_{36}E_aE_b \end{pmatrix} \quad (3)$$

Then, the SHG intensity of KDP on its *bc* plane can be expressed as:

$$I_{KDP} \propto (P^{2\omega})^2 = P_a^2 + P_b^2 + P_c^2 = (2d_{14}\sin 2\varphi)^2 \quad (4)$$

By using the *exciting* package [57], the nonlinear optical susceptibility tensor $d$ can be calculated. The calculated values of $d_{ij}$'s for $DMAGeI_3$ and KDP are compared in Supplementary Fig. 20a. Note that our calculated $d_{36}$=0.40 pm/V of KDP is close to its experimental value 0.38 pm/V[58], implying the reliability of calculation. The derived SHG anisotropy of the $DMAGeI_3$ also agrees well with the experimental measurement as presented in Supplementary Fig. 20b.

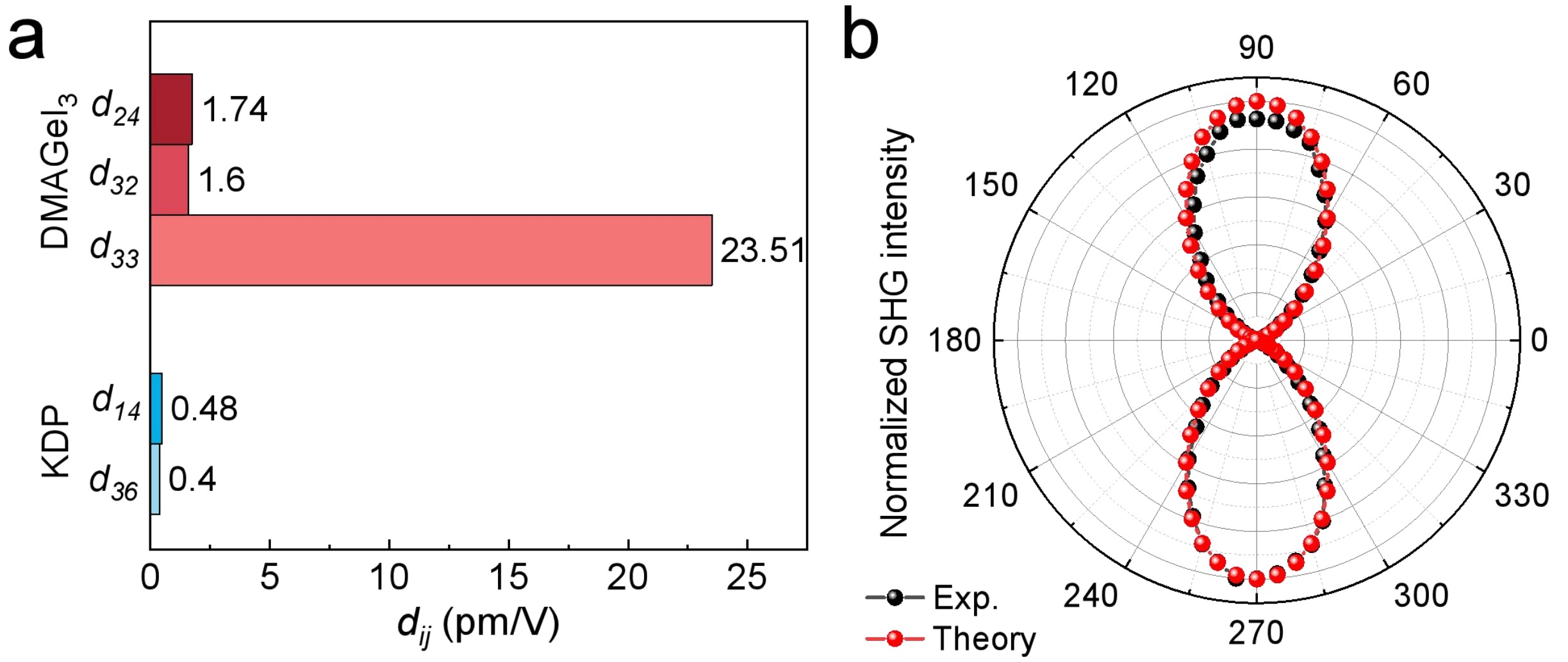

**Supplementary Fig. 20 | SHG analysis. a**, The calculated values of main $d_{ij}$'s for $DMAGeI_3$ and KDP. **b**, The normalized SHG intensity anisotropy obtained from the experimental measurement (Exp.) and our calculation (Theory) for $DMAGeI_3$.

## Supplementary References